\documentclass[a4paper,fleqn,usenatbib]{mnras}
\usepackage[T1]{fontenc}
\usepackage{ae,aecompl}

\usepackage{graphicx}	
\usepackage{amsmath}	
\usepackage{amssymb}	
\usepackage{subfig}
\usepackage{hyperref}

\usepackage{adjustbox}
\usepackage{color}
\usepackage{upgreek}
\usepackage{float}

\usepackage{booktabs}
\title[Flaring of \mbox{PKS\,0903-57}]{Unveiling the broadband spectral and temporal properties of \mbox{PKS\,0903-57} during its brightest flare}

\author[Shah et al.]{
Zahir Shah$^1$\thanks{E-mail: shahzahir4@gmail.com}, V. Jithesh$^1$, S. Sahayanathan$^2$ and Naseer Iqbal$^3$
\\
$^{1}$ Inter-University Centre for Astronomy and Astrophysics, P B No. 4, Ganeshkhind, Pune-411007, India\\
$^{2}$ Astrophysical Sciences Division, Bhabha Atomic Research Centre, Mumbai-400085, India\\
$^{3}$ Department of Physics, University of Kashmir, Srinagar 190006, India}

\date{Accepted XXX. Received YYY; in original form ZZZ}

\pubyear{2020}

\begin{document}
\label{firstpage}
\pagerange{\pageref{firstpage}--\pageref{lastpage}}
\maketitle

\begin{abstract}
We carried a detailed spectral and temporal study of blazar, \mbox{PKS\,0903-57} using the \emph{Fermi}-LAT and \emph{Swift}-XRT/UVOT observations, during its brightest flaring period MJD\,58931--58970. During this period, the maximum daily averaged $\gamma$-ray flux ($\rm F_{0.1-500\,GeV}$) of $\rm9.42\times10^{-6}\,ph\,cm^{-2}\,s^{-1}$  is observed on MJD\,58951.5, the highest $\gamma$-ray flux detected from \mbox{PKS\,0903-57} till now. Several high-energy (HE) photons ($>\,10$\,GeV) consistent with the source location at high probability (>\,99\%) are detected, and the $\gamma$-ray lightcurve in the active state shows multiple substructures with asymmetric profile. In order to understand the possible physical scenario responsible for the flux enhancement, we carried a detailed broadband spectral study of \mbox{PKS\,0903-57} by choosing different flux states from its active period.  Neglecting the multi-band variability in each of the selected time intervals, we could reproduce their averaged broadband SEDs with a one-zone leptonic model whose parameters were derived with a $\chi^2$-fit. We found that the broadband SED during different flux states can be reproduced by the synchrotron, synchrotron-self-Compton (SSC) and External-Compton (EC) processes. The seed photons for EC process from BLR or IR torus provide acceptable fits to the GeV spectrum in all the flux states; however, the detection of HE photons together with the equipartition condition suggest that  the EC/IR process is a more likely scenario. Further, a detailed comparison between the fit parameters shows that the flux enhancement from quiescent-state to the flaring-state is mostly related to increase in the bulk Lorentz factor of the emission region and change in the break energy of the source spectrum.
\end{abstract}



\section{Introduction}

Blazars are extreme class of radio loud Active Galactic Nuclei (AGN) with powerful relativistic jet pointing close to the line of sight of observer \citep{1995PASP..107..803U}. The close orientation of relativistic jet results in Doppler boosting of emission, which produces extreme observational properties like rapid flux variability across the electromagnetic spectrum \citep{2003ApJ...596..847B}, non-thermal continuum emission extending from radio to GeV/TeV energies \citep{1997ARA&A..35..445U} etc. 
The variability across the electromagnetic spectrum ranges from timescales of minutes (in some blazars
at high energies e.g, two minutes variability time scale is reported in Mrk 501 \citep{2007ApJ...669..862A},  see also \citet{2007ApJ...664L..71A} and \citet{2016ApJ...824L..20A} for short variability time scales) to several years.
These observation provide insight into the 
physics and structure of the emission regions. 
Blazars are broadly classified as BL\,Lac objects and Flat Spectrum Radio Quasars (FSRQ) depending on the characteristics of their optical spectrum. BL Lac objects have weak or absent emission line features while FSRQs show prominent emission line features.

The spectral energy distribution (SED) of blazars has two prominent broad components. 
The low energy component which peaks at optical/UV/X-ray energies, is known to be produced by synchrotron emission from the relativistic electrons in the jet. While the high energy component peaking at $\gamma$-ray energies is interpreted under different scenarios. The most commonly considered scenario is the inverse-Compton (IC) process. Here the low energy photons gets up-scattered to high energies giving rise to X-ray/GeV/TeV emission. The seed photon field for the IC scattering are synchrotron photons  (synchrotron self Compton (SSC): \citealp{1974ApJ...188..353J}, \citealp{1992ApJ...397L...5M}, \citealp{1993ApJ...407...65G}) or  photons  external to jet (external Compton (EC): \citealp{1992A&A...256L..27D}, \citealp{1994ApJ...421..153S}, \citealp{2000ApJ...545..107B}, \citealp{2017MNRAS.470.3283S}). The source of external seed photons is closely related to the location of the emission region. Alternatively, the high energy emission in blazars can be also produced by relativistic protons via hadronic processes, such as the proton synchrotron process or pion production processes \citep{1992A&A...253L..21M, 1993A&A...269...67M, 2001APh....15..121M, 2003APh....18..593M}. However, due to large jet kinetic energy required by hadronic processes, the leptonic models are some times favoured over hadronic models \citep{2013ApJ...768...54B, 2016ApJ...825L..11P}.
The location of the peak of synchrotron component in $\nu-\nu F\nu$ plot, is important observational distinction between the different classes of blazar family \citep{1998MNRAS.299..433F}. In case of FSRQs, the peak frequency of synchrotron component is located between $10^{12.5}$ and $10^{14.5}$ Hz, while the peak frequency lies between $10^{13}$ and $10^{17}$ Hz in BL\,Lac objects \citep{2010ApJ...716...30A}. The BL\,Lac objects are further classified as: Low energy peaked BL Lac objects (LBLs; $\nu_{p,syn}<10^{14}$ Hz), Intermediate energy peaked BL Lac objects (IBLs; $10^{14}$Hz $<\nu_{p,syn}<10^{15}$ Hz) and High energy peaked BL Lac objects (HBLs; $\nu_{p,syn}>10^{15}$ Hz) \citep{2010ApJ...716...30A}.

\mbox{PKS\,0903-57} is classified as blazar candidate of uncertain type (BCU) in the fourth \emph{Fermi}-LAT catalog \citep[4FGL;][]{2020ApJS..247...33A}. Importantly, the presence of  source in the first \emph{Fermi}-LAT catalog \citep[1FGL;][]{2010ApJS..188..405A} shows that \mbox{PKS\,0903-57} is bright $\gamma$-ray source.
The source is located at redshift of $z\sim0.6956$ \citep{1990PASP..102.1235T} with coordinates R.A.: 136.221579 deg, Dec.: -57.58494 deg \citep{2004AJ....128.2593F}. \mbox{PKS\,0903-57} has been reported in active state by number of Astronomy telegrams, however, detailed study of the source has not been carried out so far. For example, the daily averaged $\gamma$-ray flux of \mbox{PKS\,0903-57} in 2015 June flaring was reported 30 times greater than the average flux of the source in 3FGL catalog \citep{2015ATel.7704....1C}. The source was second time reported in high flux state by \emph{Fermi}-LAT Collaboration in May 2018, the preliminary results showed a daily averaged $\gamma$-ray flux of ${\rm 2.2\times 10^{-6}\, ph\, cm^{-2} s^{-1}}$, which is 55 times higher than its value in 3FGL catalog \citep{2018ATel11644....1C}. Recently in March--April 2020, \emph{Fermi}-LAT and AGILE observations had shown intense $\gamma$-ray activity from \mbox{PKS\,0903-57}, the activity is reported in several Astronomical telegrams \citep{2020ATel13599....1M, 2020ATel13604....1B, 2020ATel13602....1L}. Based on the preliminary analysis of the \emph{Fermi}-LAT data, a record highest $\gamma$-ray flux of ${\rm 8.2\times10^{-6}\,ph\,cm^{-2}s^{-2}}$ was reported from \mbox{PKS\,0903-57} \citep{2020ATel13604....1B}. This is the highest $\gamma$-ray flux ever observed from the source. Though \mbox{PKS\,0903-57} was absent in the first \emph{Fermi}-LAT catalog above 10 GeV \citep[1FHL;][]{2013ApJS..209...34A}, its inclusion in the third catalog of hard \emph{Fermi}-LAT sources \citep[3FHL;][]{2017ApJS..232...18A} indicates that it is not rare for this source to emit photons above 10 GeV during flaring activity. Interestingly, \emph{Fermi}-LAT observations on 2020 March 31 detected several photons above 10 GeV including a Very High Energy (VHE) photon ($\sim$ 106 GeV) consistent with the source location at high probability \citep{2020ATel13604....1B}. Later, H.E.S.S observations on 2020 April 13 resulted in a significant VHE detection ($ > 25 \sigma$) from \mbox{PKS\,0903-57} \citep{2020ATel13632....1W}. These observations provide the first evidence of VHE detections from \mbox{PKS\,0903-57}.
Following the \emph{Fermi}-LAT and \emph{AGILE} observation of intense $\gamma$-ray flaring, the \emph{Australia Telescope Compact Array} (\emph{ATCA}) observed it in radio bands on 2020 April 02. The radio fluxes reported in all bands are the highest recorded radio fluxes from the source \citep{2020ATel13638....1S}. In this work, we carried out a detailed multi-wavelength study of \mbox{PKS\,0903-57} during its active period between MJD 58880--58970 for the first time. The multi-wavelength data used in this work is acquired from \emph{Fermi}-LAT, \emph{Swift}-XRT and \emph{Swift}-UVOT. The framework of this paper is as following:- the details of the multi-wavelength data and the data analysis procedure are given in section \S\ref{sec:analy}. We presents the results of  multi-wavelength temporal and spectral analysis in section \S\ref{sec:results}. We summarise and discuss the results in section \S\ref{sec:discus}. A cosmology with $\rm \Omega_M = 0.3$, $\Omega_\Lambda = 0.7$ and $\rm H_0 = 71\rm km s^{-1} Mpc^{-1}$ is used in this work.

\section{Data Analysis}\label{sec:analy}
In order to study the multi-wavelength temporal and spectral behaviour of \mbox{PKS\,0903-57} during its flaring activity, we used the UV-Optical data from \emph{Swift}-UVOT, X-ray data from \emph{Swift}-XRT and $\gamma$-ray data from \emph{Fermi}-LAT. The details of the analysis procedure of these data sets are given below.

\subsection{\emph{Fermi}-LAT Analysis}
\emph{Fermi}-LAT principally operated in all-sky scanning mode, is a pair conversion detector sensitive to photons with energy between 20 MeV--500 GeV \citep{2009ApJ...697.1071A}. The \emph{Fermi}-LAT data for all the sources are publicly available to the scientific community usually with in 24 hours after observation.
In this work, the $\gamma$-ray data of \mbox{PKS\,0903-57} is obtained from \emph{Fermi}-LAT for the period MJD 58880--58970. The data is analysed using the \emph{Fermitools} (formally Science Tools) with version 1.2.1, these tools are hosted on an Anaconda Cloud channel that is maintained by the Fermi team. The analysis is carried by following the standard analysis procedure as described in the \emph{Fermi}-LAT documentation\footnote{http://fermi.gsfc.nasa.gov/ssc/data/analysis/}. The events were extracted in the energy range \mbox{0.1--500 GeV} from the region of interest (ROI) of 15 degree centred at the location of \mbox{PKS\,0903-57}, and using the SOURCE class events. The good time intervals (gti) are obtained by using a filter expression ''($DATA_-QUAL>0)\&\&(LAT_-CONFIG==1)$''. In order to avoid the contamination by the $\gamma$-rays from bright Earth limb, a zenith angle cut of 90 degree is applied to the data.
The Galactic diffuse emission component and the isotropic emission components were
modelled with \emph{$P8R3_-SOURCE_-V2$} and \emph{$iso_-P8R3_-SOURCE_-V2_-v1.txt$}, respectively, and the post launch instrument response function \emph{$P8R3_-SOURCE_-V2$} is used. 
In the XML model file, we included all the sources from the 4FGL catalog which were within  (15+10) degree ROI centred at the \mbox{PKS\,0903-57} location. In order to simplify the model, we initially carried out the likelihood analysis for the time period 58880--58970, the spectral parameters of sources within the 15 degree ROI were left free while parameters of sources lying outside the 15 degree ROI were fixed to their 4FGL catalogue value. Also, the photon index and normalisation of the Galactic diffuse component, and normalisation of isotropic were allowed to vary during the spectral fitting. In the XML model file, we freezed the model parameters of the background sources to their 4FGL catalog values for which $TS < 25$, and finally used the updated model for the generation of lightcurve and the spectral analysis.
We considered the detection of the source only if TS > 9 \citep[$\sim 3\sigma$ detection;][]{1996ApJ...461..396M}. Also, we used the \emph{gtsrcprob} tool and the updated XML model to calculate the probability of HE event ($>$10 GeV) of being originated from the direction of \mbox{PKS\,0903-57}.

\subsection{\emph{Swift} Analysis}
The X-ray data in the energy range 0.3--10 keV from XRT on board the \emph{Neil Gehrels Swift Observatory} \citep[\emph{Swift};][]{2004ApJ...611.1005G} has been used in this work. The small field of view of \emph{Swift} allows the pointing mode observations only which are carried either in monitoring program or target of opportunity (ToO) program. The data of \mbox{PKS\,0903-57} in our work was obtained as ToO observations (PIs: Mereu and Wierzcholska). During the period MJD 58880--58970, a total of 12 \emph{Swift} observations of \mbox{PKS\,0903-57} are available. The total exposure time of these observations are approximately 21.2 ks. The \emph{Swift}/XRT lightcurve is obtained such that each point in the lightcurve corresponds to one observation ID. The X-ray data obtained in photon-counting (PC) mode is processed with the XRTDAS V3.0.0 software package and the Standard xrtpipeline (Version: 0.13.4) is used to create the cleaned event files. The input parameters to the \emph{xrtpipeline} is given by following the \emph{Swift} X-ray data analysis thread page. Depending on the count rate, the source and background regions are chosen by using the \emph{xrtgrblc} (v1.9) task \citep{2013ApJS..207...28S}. The source region is chosen as circular region if the count rate is $\leq 0.5~\rm counts~s^{-1}$ and annular region if count rate is $>0.5~\rm counts~s^{-1}$. While background region is chosen as annular region in all cases. 
Further, we used an automated products generator tool available at the UK \emph{Swift} Science Data Centre \citep{2009MNRAS.397.1177E} to obtain the data products (source, background, and ancillary response files) for spectral analysis in different flux states. In order to make the spectrum valid for the $\chi^2$-statistics, the GRPPHA task is used to rebin the source spectra so that the resultant spectra have 20 counts per bin. The X-ray spectral analysis in the energy range 0.3--10 keV was performed using the XSPEC package \citep{1996ASPC..101...17A} builded in HEASOFT. The 0.3--10 keV spectra were fitted with an absorbed power law by fixing the Galactic neutral hydrogen column density $N_H= 2.6\times 10^{21}~\rm cm^{-2}$ while keeping the normalisation and the spectral index as free parameters.

\subsection{\emph{Swift}-UVOT}
The \emph{Swift}-UVOT \citep{2005SSRv..120...95R} observed the \mbox{PKS\,0903-57} in optical and UV with the filters v, b, u; and w1, m2, and w2 \citep{2008MNRAS.383..627P, 2010MNRAS.406.1687B}. The data acquired from the HEASARC Archive and were analysed with the \emph{uvotsource} task included in the HEASoft package (v6.26.1). We have used the most recent Calibration files of UVOT (as of 15 December 2020) for the analysis. In case of multiple images in the filter, we added the images using the \emph{uvotimsum} tool. Source counts were
extracted from a circular region of radius 5$''$ centred at the source location, whereas background counts were extracted from a nearby source free circular region of radius 10$''$. The observed fluxes were de-reddened for Galactic extinction using $\rm E(B-V)=0.329$ and $R_{V} = A_{V}/E(B-V)=3.1$ following \citet{2007ApJ...663..320F}.

\section{Results}\label{sec:results}
\mbox{PKS\,0903-57} was reported in intense $\gamma$-ray activity by observations from \emph{Fermi}-LAT and \emph{AGILE} during the time period MJD 58931--58970 \citep{2020ATel13599....1M, 2020ATel13604....1B, 2020ATel13602....1L}. Based on the preliminary analysis of the \emph{Fermi}-LAT data, a record maximum $\gamma$-ray flux of $\rm 8.2\times 10^{-6}\rm~ph~cm^{-2}~s^{-1}$ was reported from \mbox{PKS\,0903-57} \citep{2020ATel13604....1B}. The reported flux is the highest $\gamma$-ray flux ever observed from this source. Also, several high energy photons ($> 10$ GeV) with high probability of being associated with the source location are detected during this period \citep{2020ATel13604....1B}. Moreover, H.E.S.S. has made a significant VHE detection ($> 25$ standard deviations) from \mbox{PKS\,0903-57} on April 13, 2020 for a total observation time of 49 minutes \citep{2020ATel13632....1W}.
Motivated by the detection of highest $\gamma$-ray flux, observation of VHE photons and the availability of multi-wavelength observations, we carried a detailed multi-wavelength study of \mbox{PKS\,0903-57} using the \emph{Fermi}-LAT and \emph{Swift}-XRT/UVOT observations. The aim is to understand the temporal and spectral characteristics of source. Such study has never been carried on this source.

\subsection{Temporal Study}
\begin{figure*}
		\begin{center}
        \includegraphics[scale=0.5]{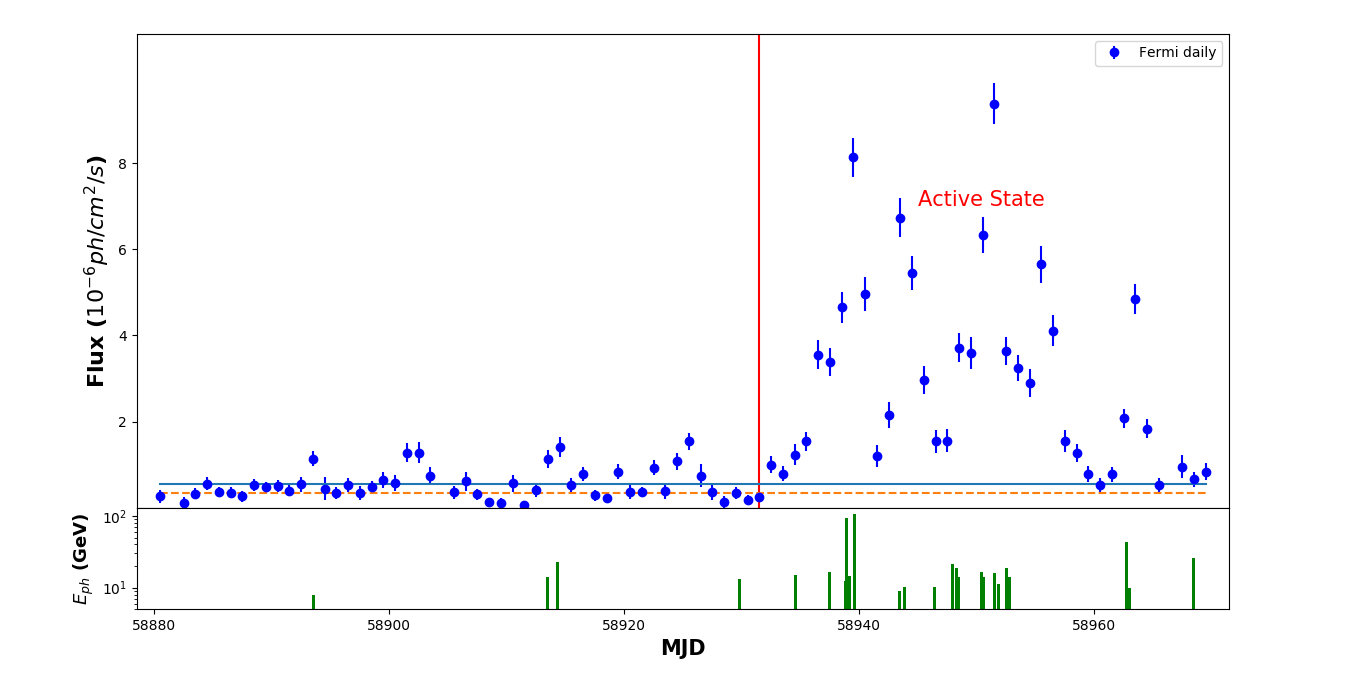}
        \caption{Top panel: One-day binned \emph{Fermi}-LAT $\gamma$- ray lightcurve of \mbox{PKS\,0903-57} obtained by integrating over the energy range \mbox{0.1--500 GeV} during the period MJD 58880--58970. Bottom panel: The energy of observed HE photons with 99\% or higher probability of being associated with \mbox{PKS\,0903-57},  shown by green bars.}       
        \label{fig:gamma_lc}
		\end{center}        
        \end{figure*}

\begin{figure*}
		\begin{center}
        \includegraphics[scale=0.7]{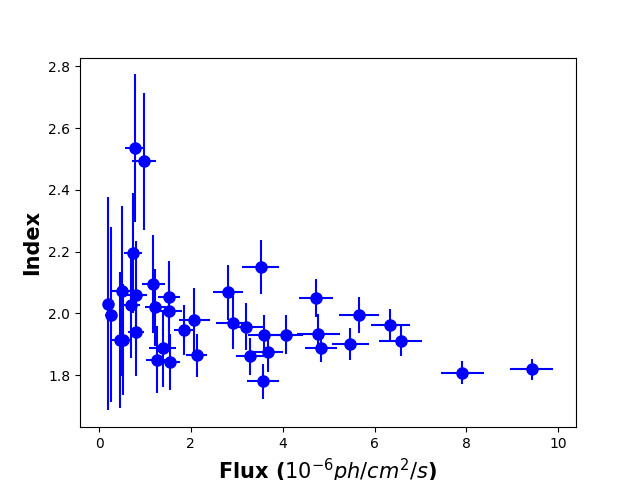}
        \caption{Variation of power-law index plotted as a function of the flux (above 100 MeV) 
        during the active state of \mbox{PKS\,0903-57}.}      
        \label{fig:index_flux}
		\end{center}        
        \end{figure*}

In the top panel of Figure \ref{fig:gamma_lc}, we present the daily binned $\gamma$-ray lightcurve obtained during the period MJD 58880--58960. 
As shown in figure, the source started a major $\gamma$-ray activity around MJD 58931 (time period marked after solid vertical line with red colour in Figure \ref{fig:gamma_lc}). In the figure, the blue horizontal line depicts the average value of fluxes before the start of major activity (< MJD 58931), it represents a flux level of $\rm 5.50\times 10^{-7}\rm~ph~cm^{-2}~s^{-1}$. Here we obtain the base flux $\rm F_0=3.33\times 10^{-7}\rm~ph~cm^{-2}~s^{-1}$ as the average value of the fluxes which are below $\rm 5.50\times 10^{-7}\rm~ph~cm^{-2}~s^{-1}$, it is represented by  orange dashed horizontal line in Figure \ref{fig:gamma_lc}. As shown in figure, the source flux remained above the base line flux for more than a month during the time period 2020 March 23--2020 May 01 (MJD 58931--58970).
Considering the long-term activity, we defined this time period 2020 March 23--2020 May 01 as an active state of \mbox{PKS\,0903-57}, and carried a detailed temporal and spectral characteristics of the source during this period. The daily binned $\gamma$-ray lightcurve in the active state displays five peaks with a maximum  flux of $(9.42\pm0.47)\times10^{-6}\rm~ph~cm^{-2}~s^{-1}$ corresponding to spectral index of $1.82\pm0.03$ observed on MJD 58951.5. This is the highest daily binned $\gamma$-ray flux detected from this source till now and is factor of $\sim 28$ larger than the base flux $F_0$.  The $\gamma$-ray lightcurve shown in Figure \ref{fig:gamma_lc} is obtained by fitting the daily binned integrated spectrum (100 MeV--500 GeV) with a power law model, $\frac{dN}{dE}=N_0(E/E_0)^{\Gamma_i}$, where $N_0$ is normalisation, $E_0$ is scale energy, $\Gamma_i$ is power law index. The plot between the daily binned $\gamma$-ray flux and the spectral
index (shown in Figure \ref{fig:index_flux}) suggests that the source exhibits a harder
when brighter trend.
 We used the Spearman rank correlation method to statistically quantify the correlation between the flux and index. The obtained correlation coefficient, $\rho$=-0.50 and null-hypothesis probability value, $P_{value}=2.66\times10^{-3}$ further confirms a harder when brighter trend in \mbox{PKS\,0903-57}, a common feature identified in blazars \citep[e.g][]{2016ApJ...830..162B, 2019MNRAS.484.3168S}. 
Since the $Fermi$-LAT energy lies near the peak of IC component of SED, spectral hardening during
the flaring state indicates the detection of more number of high
energy photons. This may happen when  the IC peak of the
SED shift to higher energy during high flux state. Incidentally such feature with a  strong correlation between spectral
hardening during flare and shift in SED peak energy of IC component is reported by \citet{2019MNRAS.484.3168S}  in the 2018 January flare of 3C\,279.

\begin{figure*}
		\begin{center}
	\includegraphics[width=20cm,height=12cm]{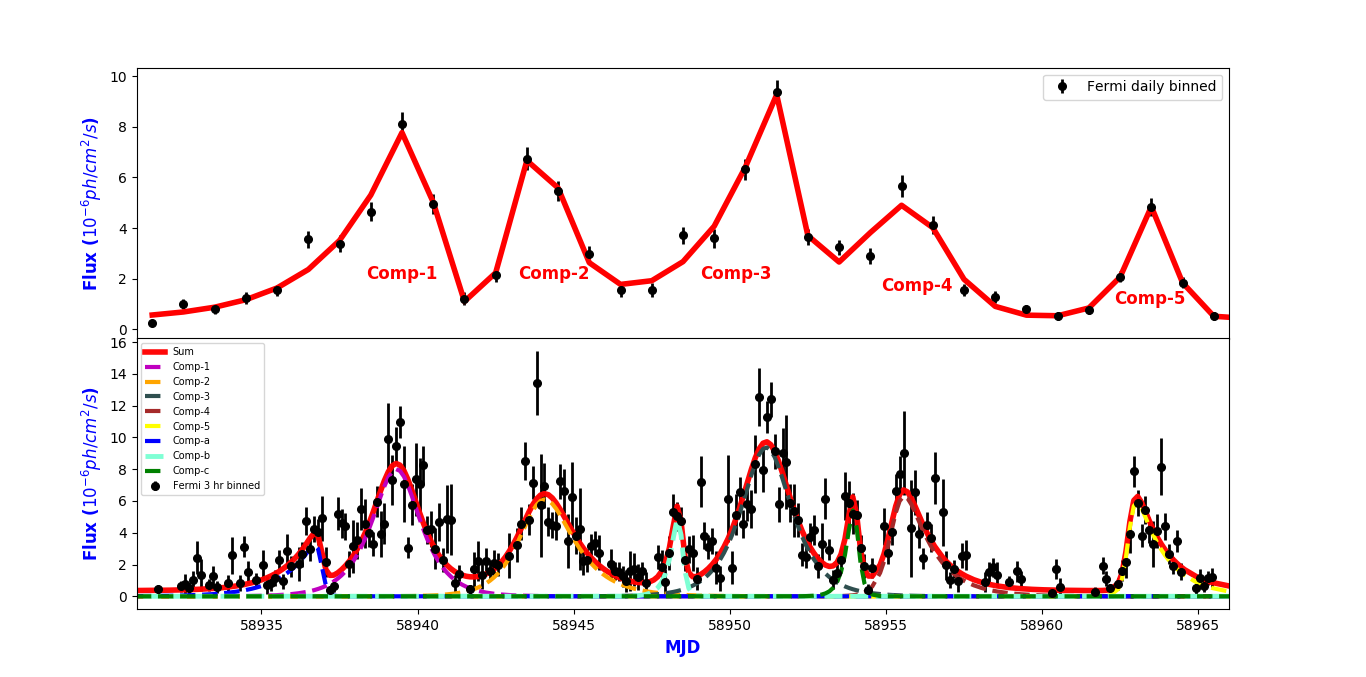}
        \caption{Top panel: Daily binned $\gamma$-ray lightcurve of \mbox{PKS\,0903-57} obtained in the active state (MJD 58931--58970) by integrating over the energy range \mbox{0.1--500 GeV}. The solid red curve is the best fitted sum of exponential function.
Bottom panel: 3-hour binned $\gamma$-ray (\mbox{0.1--500 GeV}) lightcurve obtained in the active state which is used to find the shortest variability time scale $t_{var}$. Again the solid red curve is the best fitted sum of exponential function while the dashed curves with different colours are the individual component fits. }    
        \label{fig:daily_3h_rise_fal}
		\end{center}        
        \end{figure*}

Considering the good photon statistics at $\gamma$-ray energy in the active state, we obtained a three hour binned $\gamma$-ray lightcurve and same is shown in the second panel of Figure \ref{fig:daily_3h_rise_fal}. This three hour binned $\gamma$-ray lightcurve shows a peak flux of $\rm (1.34\pm0.02)\times 10^{-5}\rm~ph~cm^{-2}~s^{-1}$ with a photon index of $2.00\pm0.10$ on MJD 58943.81. The 
three hour binned $\gamma$-ray lightcurve displays three additional structures (Comp-a, Comp-b and Comp-c) in addition to the five dominant components (Comp-1, Comp-2, Comp-3, Comp-4 and Comp-5) which are present in the daily binned $\gamma$-ray lightcurve (see top and bottom panel of Figure \ref{fig:daily_3h_rise_fal}). In order to obtain the estimates of the rise and decay times of these components, we fitted the three hour binned $\gamma$-ray lightcurve with the expression
\begin{equation}\label{eq:rise_fall}
F(t)=F_b+\Sigma_{i\rightarrow 1}^{8}F_{i}(t)\quad,
\end{equation}
where $F_b$ is the base line flux and
\begin{equation}
F_{i}(t)=\frac{2F_{p,i}}{\exp\left(\frac{t_{p,i}-t}{\tau_{r,i}}\right)+\exp\left(\frac{t-t_{p,i}}{\tau_{d,i}}\right)} \quad, \nonumber
\end{equation}
Here, $F_{p,i}$ is peak flare amplitude at peak time $t_p$, $\tau_{r,i}$ and $\tau_{d,i}$ are the rise and decay times of the respective flare component. The fitted profile along with multiple components of active state is shown in the bottom panel of Figure \ref{fig:daily_3h_rise_fal} and the best-fit parameter values are given in Table \ref{table:multi-comp}. We measured the strength of asymmetry of the flaring components by using the parameter $\zeta=\frac{\tau_d-\tau_r}{\tau_d+\tau_r}$, such that the flare component will be symmetric, if $|\zeta|<0.3$; moderately asymmetric flares, if $0.3 <|\zeta| < 0.7$; and markedly asymmetric flares, if $0.7 < |\zeta| < 1$ \citep{2010ApJ...722..520A}. As shown in Table \ref{table:multi-comp}, the Comp-5 and Comp-a show a considerable asymmetry with $\zeta=$ 0.76 and 0.84, respectively, while Comp-4, Comp-b and Comp-c with $\zeta$=0.54,  0.43, and 0.31 respectively shows moderate asymmetry.
Further, in order to obtain an estimate of shortest variability time scale, we scanned the three hour binned $\gamma$-ray lightcurve with the equation
\begin{equation}
F(t)=F(t_0) 2^{\frac{t-t_0}{\tau}}\quad ,
\end{equation}
where $F(t)$ and $F(t_0)$ are the flux at time t and $t_0$, respectively and $\tau$ is characteristic doubling time scale.
Using the condition that the significance of difference in flux at $t$ and $t_0$ is $\geq3\sigma$ \citep{2011A&A...530A..77F}, the shortest flux doubling timescale $t_{var}=0.94$ hr is obtained in the Comp-1 at MJD 58937.3. 
We also searched for the most energetic $\gamma$-ray photons from \mbox{PKS\,0903-57} using the \emph{gtsrcprob} tool and event class CLEAN. The \emph{gtsrcprob} tool calculates probability with which a photon can be associated with particular source in the XML model. A closer inspection of the individual events revealed presence of several HE photons positionally coincident with the \mbox{PKS\,0903-57} location. In the bottom panel of Figure \ref{fig:gamma_lc}, the observed HE photons (>10 GeV) with 99\% or higher probability of being associated with the source are plotted (green bars) against the time of detection. The $\gamma$-ray photon with highest energy of 106 GeV is detected on MJD 58939.58088 (2020 April 10 02:16:12.689) with 99.78 per cent probability that it originated from \mbox{PKS\,0903-57}.

\begin{table*}
\caption{Multiple components in the active state fitted with sum of exponential functions (equation \ref{eq:rise_fall}) in order to obtain rise and fall time of the individual components. Col. 1: individual components; 2, 3 and 4: peak time, rise time and decay time of the components; 5: asymmetry parameter. The reduced-$\chi^2$ is obtained as 2.54.}
\begin{adjustbox}{width=0.5\textwidth,center=\textwidth}
\begin{tabular}{lcccr}

\bottomrule
Component  & $t_p$ &  $t_{r}$ (days) & $t_{d}$ (days) & $|\zeta|$ \\
\bottomrule

Comp-1 & 58939.33$\pm$0.25 & 0.71$\pm$0.19 & 0.68$\pm$0.16 & 0.02 \\
Comp-2 & 58943.93$\pm$0.33 & 0.65$\pm$0.22 & 0.93$\pm$0.23 & 0.18\\
Comp-3 & 58951.14$\pm$0.26 & 0.70$\pm$0.17 & 0.75$\pm$0.18 & 0.03\\
Comp-4 & 58955.33$\pm$0.17 & 0.27$\pm$0.13 & 0.90$\pm$0.21 & 0.54\\
Comp-5 & 58962.80$\pm$0.09 & 0.13$\pm$0.05 & 0.94$\pm$0.17 & 0.76 \\
Comp-a & 58936.90$\pm$0.13 & 1.01$\pm$0.27 & 0.09$\pm$0.08 & 0.84 \\
Comp-b & 58948.40$\pm$0.12 & 0.25$\pm$0.13 & 0.10$\pm$0.07 & 0.43 \\
Comp-c & 58954.00$\pm$0.15 & 0.23$\pm$0.13 & 0.12$\pm$0.09 & 0.31 \\

\bottomrule

\end{tabular}
\end{adjustbox}

\label{table:multi-comp}
\end{table*}

\begin{figure*}
		\begin{center}
        \includegraphics[scale=0.5]{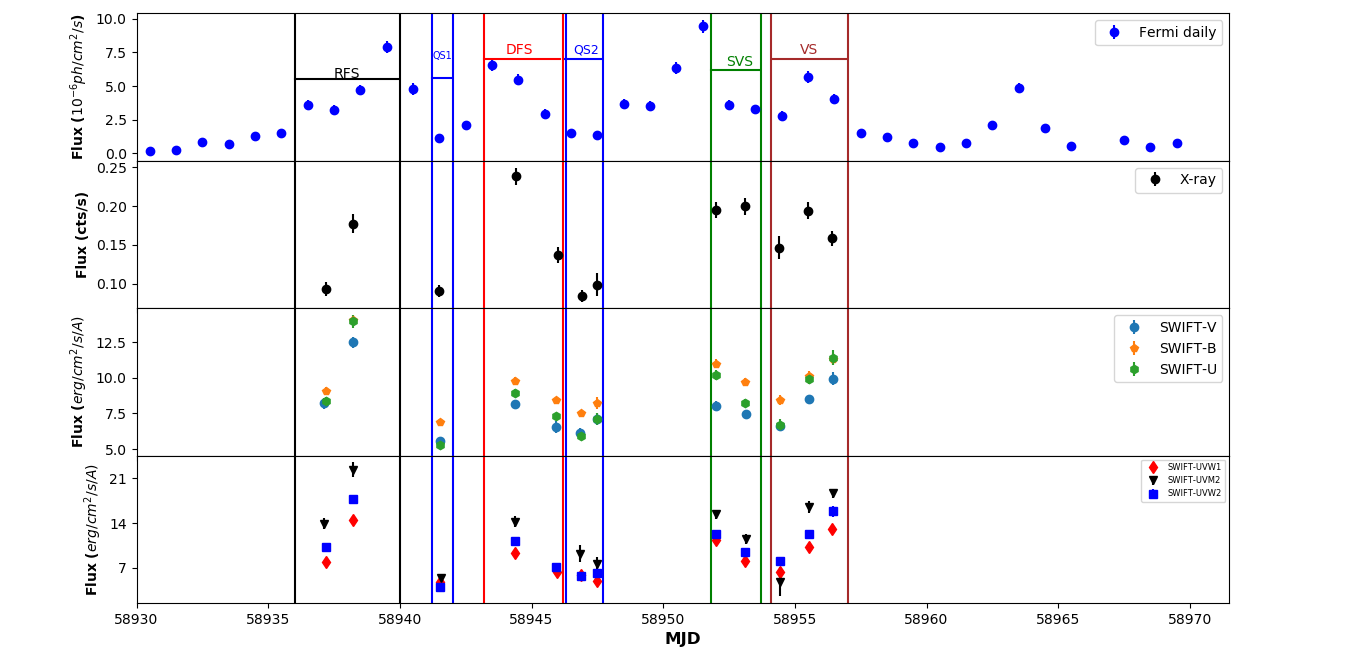}
        \caption{ Multi-wavelength lightcurves (MLCs) of \mbox{PKS\,0903-57} obtained by using \emph{Fermi}-LAT and \emph{Swift} XRT and UVOT observations during the active state (defined in Figure \ref{fig:gamma_lc}). Top panel is daily binned $\gamma$-ray lightcurve obtained by using \emph{Fermi}-LAT data in the energy range \mbox{0.1--500 GeV}. The second panel is the X-ray lightcurve obtained by using the \emph{Swift}-XRT data in the energy range 0.3--10 keV. Third and fourth panel are UV and optical lightcurves with fluxes in units of $\rm 10^{-15}$ erg/$cm^2$/s/{\AA} obtained using \emph{Swift}-UVOT data. The regions represented with the coloured vertical lines define different flux states over which broadband spectral behaviour of the source is studied in detail.}       
        \label{fig:multiplot_lc}
		\end{center}        
        \end{figure*}

Following the high activity at $\gamma$-rays, \emph{Swift}-XRT and UVOT also carried a total of 12 observations of \mbox{PKS\,0903-57}, to unveil the behaviour of source at X-ray, UV and optical bands. Figure \ref{fig:multiplot_lc} shows the multi-wavelength lightcurve of \mbox{PKS\,0903-57} obtained during the active state, using the observations from \emph{Fermi}-LAT, \emph{Swift}-XRT and UVOT. In this multiplot, the top panel shows daily binned $\gamma$-ray lightcurve with flux points obtained by integrating over the energy range \mbox{0.1--500 GeV}; the X-rays, UV and optical lightcurves are shown in the upper middle, lower middle and bottom panels, respectively such that each flux point corresponds to individual observation ID. The multiplot lightcurves give the impression of correlated flux variations in different energy bands. Considering the $\gamma$-ray lightcurve as a reference lightcurve, we used the Spearman rank correlation method to statistically quantify the correlation between different energy bands. The $\rho$ and $P_{value}$ obtained between lightcurves (see Table \ref{table:corr_mlc}) further confirm the correlated flux variations between different energy bands.
\begin{table*}
\caption{Correlation between the $\gamma$-ray lightcurve with the X-ray and optical/UV lightcurves using Spearman rank correlation method. Col. 1: Lightcurves between correlation is calculated, 2: Correlation coefficient value 3: Probability value of null-hypothesis.}
\begin{adjustbox}{width=0.4\textwidth,center=\textwidth}
\begin{tabular}{lcr}
\bottomrule
Lightcurves & $\rho$  & $P_{value}$   \\
\bottomrule
$\gamma$-ray vs X-ray & 0.83 &  $9.51\times10^{-4}$ \\
$\gamma$-ray vs V     &  0.74 &  $4.51\times10^{-3}$ \\
$\gamma$-ray vs B     &  0.89 &   $1.11\times10^{-4}$ \\
$\gamma$-ray vs U    &  0.87 &   $2.60\times10^{-4}$\\
$\gamma$-ray vs W1  &  0.90 &   $8.36\times10^{-5}$\\
$\gamma$-ray vs M2  &  0.79  &   $3.74\times10^{-3}$ \\
$\gamma$-ray vs W2  & 0.87  &   $2.59\times10^{-4}$ \\
\bottomrule
\end{tabular}
\end{adjustbox}

\label{table:corr_mlc}
\end{table*}
To quantify the variability amplitude of \mbox{PKS\,0903-57} in different energy bands, we calculated the fractional variability amplitude using the expression \citep{2003MNRAS.345.1271V}
\begin{equation}\label{eq:fvar}
F_{var}=\sqrt{\frac{S^2-\overline{\sigma_{err}^2}}{\overline{F}^2}}
\end{equation}
where $S^2$ and $\overline{F}$ are the lightcurve variance and mean, and $\overline{\sigma_{err}^2}$ is the mean square of the measurement error. The uncertainty on $F_{var}$ is  given by \citep{2003MNRAS.345.1271V}
\begin{equation}
F_{var,err}=\sqrt{\frac{1}{2N}\left(\frac{\overline{\sigma_{err}^2}}{F_{var}\overline{F}^2}\right)^2+\frac{1}{N}\frac{\overline{\sigma_{err}^2}}{\overline{F}^2}}
\end{equation}

\begin{figure*}
		\begin{center}
        \includegraphics[scale=0.7]{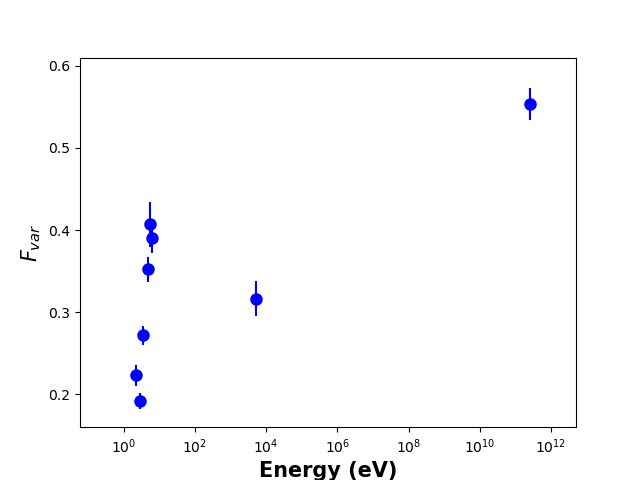}
        \caption{Fractional variability obtained in optical, UV, X-ray and $\gamma$-ray energy bands plotted as function of their energy. The $F_{var}$ values are reported in Table \ref{table:var_amp}.}       
        \label{fig:frac_var}
		\end{center}        
        \end{figure*}

where N is the number of flux points in the lightcurve. The $F_{var}$ values obtained in the considered energy bands are given in Table \ref{table:var_amp}. 
The  $F_{var}$ plotted as function of energy in Figure 5 shows 
an increasing trend in the optical/UV filters (with some fluctuations), the increase in $F_{var}$ is by a factor of $\sim 2$ from optical to UV filter. Following the peak in optical/UV band, the $F_{var}$ shows a dip in the X-ray band and  then it again increases in the $\gamma$-ray band. The dip in $F_{var}$ vs energy plot has been
reported in several blazar works \citep[e.g,][]{2016ApJ...819..156B, 2016A&A...590A..61C, 2017MNRAS.464..418R}.
In Figure \ref{fig:frac_var}, the dip could be related to the double hump feature in the broadband SED of
blazars such that the highest energy electrons dominate the emission at UV and GeV energies through synchrotron and inverse Compton processes respectively; whereas, the emission at keV energies could be of inverse Compton origin by relatively low energy electrons. Therefore, the variability can be manifestation of cooling of relativistic electron such that higher energy electrons cool faster than the low energy electrons, which results in faster variability in the emission from the high energy electrons. 

 The $F_{var}$ values show significant change in the optical/UV filters. Though we have de-reddened the observed fluxes for Galactic extinction, it is possible that a systematic error exists, which may account for large variability changes observed in the closely spaced optical/UV filters.  Also, it should be noted that the obtained $F_{var}$ values do not include the timing information of the lightcurves.  The gaps in the lightcurve (see Figure \ref{fig:multiplot_lc}) can also effect the $F_{var}$ results, for example \citet{2019Galax...7...62S} showed that the spread in the fractional variability increases with the removal of time bins, while it decreases with the completeness of the lightcurve.

\begin{table*}
\caption{The fractional  variability amplitude ($F_{var}$) of \mbox{PKS\,0903-57}, calculated using equation \ref{eq:fvar} in different energy bands during the time period MJD 58935--58959}
\begin{adjustbox}{width=0.35\textwidth,center=\textwidth}
\begin{tabular}{lr}
\bottomrule
Energy band  & $F_{var}$ \\
\bottomrule
$\gamma$-ray (\mbox{0.1--500 GeV}) & 0.55$\pm$0.02  \\
X-ray (0.3--10 keV) & 0.32$\pm$0.02  \\
UVW2 & 0.39$\pm$0.02 \\
UVM2 & 0.41$\pm$0.03\\
UVW1 & 0.35$\pm$0.02\\
U & 0.27 $\pm$ 0.01\\
B & 0.19 $\pm$ 0.01\\
V & 0.22 $\pm$ 0.01\\
\bottomrule
\end{tabular}
\end{adjustbox}

\label{table:var_amp}
\end{table*}

\begin{table*}
\caption{Results of the integrated $\gamma$-ray spectrum of  RFS, DFS, QS1 and QS2, SVS and VS modelled with the power law and log-parabola model. Col. 1: flux state; 2: time period of flux state; 3: fitted model; 4: integrated flux in$\rm~ph~cm^{-2}~s^{-1}$; 5: power law index or index defined at reference energy; 6: curvature parameter; 7: test statistics; 8: -log(likelihood); 9: significance of curvature.}
\begin{adjustbox}{width=1.0\textwidth,center=\textwidth}
\begin{tabular}{lcccccccr}
\bottomrule
State  & period & Model & $F_{0.1-500 GeV}$ & $\Gamma$ or $\alpha$ & $\beta$ & TS & -$\log\mathcal{L}$ & $TS_{curve}$ \\
\bottomrule
RFS & 58936.0--58940.0 & PL & $5.12\pm0.20$ & $1.92\pm0.03$  & ---  &  4174.05  &  13708.52 & ---  \\
 &  & LP & $4.81\pm0.21$ & $1.90\pm0.03$ & $0.07\pm0.02$ & 4146.46 & 13700.39 & 16.26\\
DFS & 58943.2--58946.2 & PL & $5.00\pm0.24$ & $1.94\pm0.03$ & ---  & 2697.72 & 10293.11 & --- \\
	&  & LP & $4.75\pm0.25$ & $1.91\pm0.04$ & $0.06\pm0.02$ & 2743.50 & 10290.23 & 5.76\\
QS1 & 58941.2--58942.0 & PL &  $0.96\pm0.27$ & $2.00\pm0.19$ & --- & 67.20 & 2823.19 & ---\\
      &  &  LP  &  $0.84\pm0.34$ & $1.94\pm0.25$  &  $0.08\pm0.14$ & 68.55 & 2821.60 & 3.18\\
QS2 & 58946.3--58947.7 & PL & $1.64\pm0.23$ & $2.01\pm0.10$ & --- & 298.53 &  4770.63 & ---\\
	& & LP & $1.57\pm0.24$  & $1.99\pm0.10$ & $0.03\pm0.06$ & 296.50 & 4773.11 & -4.95 \\
SVS & 58951.8--58953.7 & PL &  $3.5\pm0.24$ & $1.83\pm0.04$ & --- & 1453.51 & 6833.861 & ---\\
	& & LP & $3.08\pm0.25$  & $1.75\pm0.06$ & $0.12\pm0.04$ & 1445.68 & 6825.800 & 16.12 \\
VS & 58954.1--58957 & PL &  $4.44\pm0.25$ & $2.01\pm0.04$ & --- & 1907.31 & 9459.19 & ---\\
	& & LP & $4.06\pm0.25$  & $2.01\pm0.05$ & $0.11\pm0.03$ & 1951.19  & 9450.27 &  17.84\\
\bottomrule
\end{tabular}
\end{adjustbox}

\label{table:spec_fit_param}
\end{table*}

\begin{table*}
\caption{Best-fit source parameters and properties of \mbox{PKS\,0903-57} obtained by fitting the local XSPEC SED emission model \citep{2018RAA....18...35S} to the RFS, DFS, PS, QS1 and QS2. Here we consider two cases of target photon field for EC scattering viz. target photon field from IR region (with temperature 900 K) and BLR region (dominant Layman alpha line emission at frequency $\rm 2.5\times10^{15} Hz$) \citep{2018RAA....18...35S}.
Row:- 1: particle energy density in units of $\rm 10^{-3} erg\,cm^{-3}$, 2: electron Lorentz factor corresponding to the break energy, 3: bulk Lorentz factor of the emission region, 4: magnetic field in units of $\rm G$, 5: Doppler factor 6: Reduced $\chi^2$/degrees of freedom 7: logarithmic jet power in units of $\rm{erg\,s^{-1}}$, 7: logarithmic total radiated power in units of $\rm{erg\,s^{-1}}$ and 8: ratio of particle energy density and magnetic field energy density. Following parameters were kept fixed for all flux states: low energy and high energy power-law index of the particle distribution (p and q), size of emission region $R'$ (cm), minimum and maximum Lorentz factor of particle distribution $\rm \gamma_{min}$ and $\rm \gamma_{max}$, viewing angle $\theta$ and covering factor $f$.  In case the target photons are from IR region, $\rm R=5.21\times10^{16}\, cm$, $p=2.67$, $q=5.69$, $\rm \gamma_{min}=50$, $\rm \gamma_{max}=10^8$, , $\rm\theta=0.1 \,degree$ and $f=3.19\times 10^{-3}$, while as for target photon field from BLR region, $\rm R=1.88\times10^{16}\,cm$, $p=2.26$, $q=4.97$, $\rm \gamma_{min}=20$, $\rm \gamma_{max}=10^6$, $\rm\theta=0.1\,degree$ and $f=10^{-8}$. The subscript and superscript values on parameter are lower and upper values of model parameters respectively obtained through spectral fitting. $--$ implies that the upper or lower bound value on the parameter is not constrained.}
\begin{adjustbox}{width=1.05\textwidth,center=\textwidth}
\begin{tabular}{lcccccccccccccccccc}
\bottomrule
& \multicolumn{6}{c}{IR photons} && \multicolumn{6}{c}{BLR photons} \\  \cmidrule(lr){2-7}  \cmidrule(lr){9-14}
Parameters & RFS & DFS  & QS1 & QS2 & SVS & VS && RFS & DFS  & QS1 & QS2 & SVS & VS  \\
\bottomrule
\vspace{2.5mm}
$U_e$ & $4.68_{3.47}^{--}$  & $9.59_{6.93}^{--}$   & $3.91_{2.00}^{5.01}$ & $6.26_{2.71}^{7.97}$  & $5.85_{4.22}^{--}$ &  $5.64_{4.06}^{--}$ &&  $24.22^{11.24}_{--}$ & $46.96_{17.30}^{--}$ &  $16.42_{4.78}^{--}$ & $22.39_{8.63}^{--}$ & $29.92_{11.82}^{--}$ &  $38.67_{22.70}^{--}$\\
\vspace{1.5mm}
$\gamma_b$ & $4680_{3731}^{6416}$ & $5114_{4063}^{7823}$  & $2139_{1249}^{2656}$ &  $3520_{1668}^{4391}$ & $4977_{3814}^{7974}$ & $4906_{3716}^{6592}$ && $2814_{1198}^{4305}$ & $2892_{1201}^{4813}$ & $1029_{315}^{1010}$ & $1570_{--}^{2645}$ & $2926_{--}^{4851}$ & $4242_{2577}^{6596}$ \\
\vspace{1.5mm}
$\Gamma$ & $29.71_{27.50}^{31.60}$ & $26.15_{24.07}^{27.94}$   & $23.05_{20.85}^{--}$ & $23.04_{21.67}^{--}$ & $27.38_{25.03}^{29.32}$ & $25.43_{23.48}^{27.22}$ && $30.95_{28.10}^{33.16}$ & $27.22_{24.17}^{--}$ &  $23.60_{19.69}^{--}$ & $25.20_{22.14}^{--}$ & $28.25_{25.55}^{--}$ &  $27.19_{24.83}^{30.16}$\\
\vspace{1.5mm}
$B$ & $0.24_{0.19}^{0.27}$ & $0.18_{0.14}^{0.22}$  &    $0.48_{0.40}^{--}$ & $0.29_{0.26}^{--}$ & $0.22_{0.17}^{0.26}$ &  $0.27_{0.21}^{0.32}$ && $0.37_{0.26}^{0.60}$ & $0.31_{0.18}^{0.30}$ &  $1.12_{0.65}^{--}$ & $0.67_{0.42}^{--}$ & $0.35_{0.24}^{0.52}$ &  $0.27_{0.20}^{0.39}$\\
\vspace{1.5mm}
$\chi^2_{red.}$/dof & 0.82/13 & 0.78/13 & 0.64/12 &  0.74/11 & 0.81/13 & 0.78/13  &&  0.94/13 & 0.83/13 & 0.79/12 & 0.72/11 & 0.73/13 & 0.82/13\\
\hline
Properties &&& \\ 

\hline
\vspace{1.5mm}
$\delta$ & 59.24 & 52.18  & 46.00 & 45.99 & 54.62 & 50.75 && 61.70 &  54.31 & 47.11 & 50.29 &  56.35 & 54.24\\
\vspace{1.5mm}
$P_{\rm jet}$ & 46.26 & 46.45  & 46.01 & 46.17 & 46.28 & 46.20 && 46.32 & 46.49 & 46.00 & 46.14 & 46.33 & 46.39\\
\vspace{1.5mm}
$P_{\rm rad}$ & 42.14  & 42.36  & 41.87 & 42.07 & 42.18  & 42.30 && 41.98 & 42.20 &  41.79 & 41.87 & 42.03 & 42.10\\
\vspace{1.5mm}
$U_e/U_B$ & 2.13 & 7.15  & 0.42 & 1.83 & 3.11 & 2.01 && 4.57 & 12.45 & 0.33 & 1.28 & 6.13 & 13.00 \\
\hline
\end{tabular}
\end{adjustbox}

\label{table:sed}
\end{table*}

\begin{figure*}
		\begin{center}

	\includegraphics[angle=0,width=.58\textwidth]{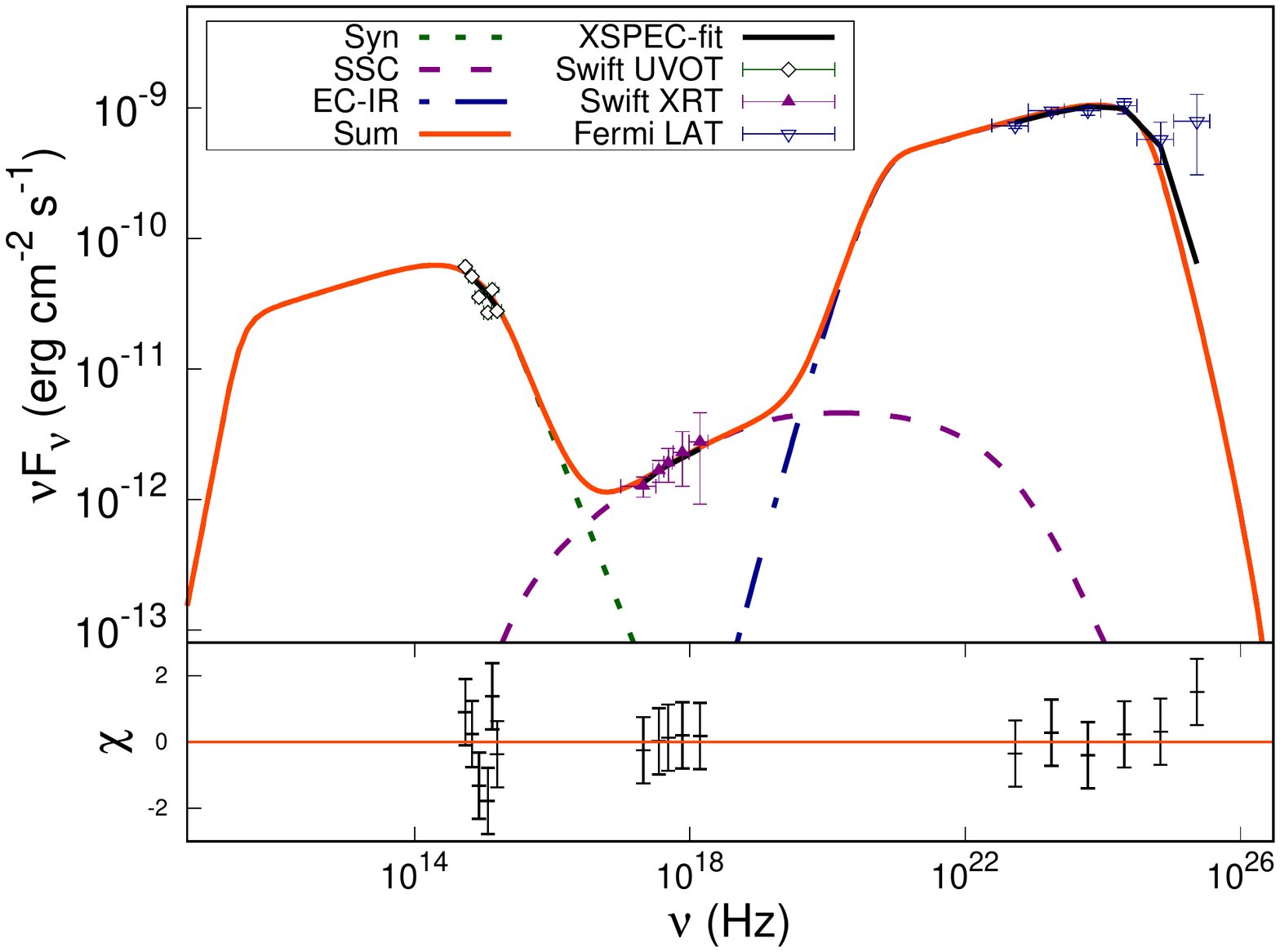}\hspace{-3.0cm}
        \includegraphics[angle=0,width=.58\textwidth]{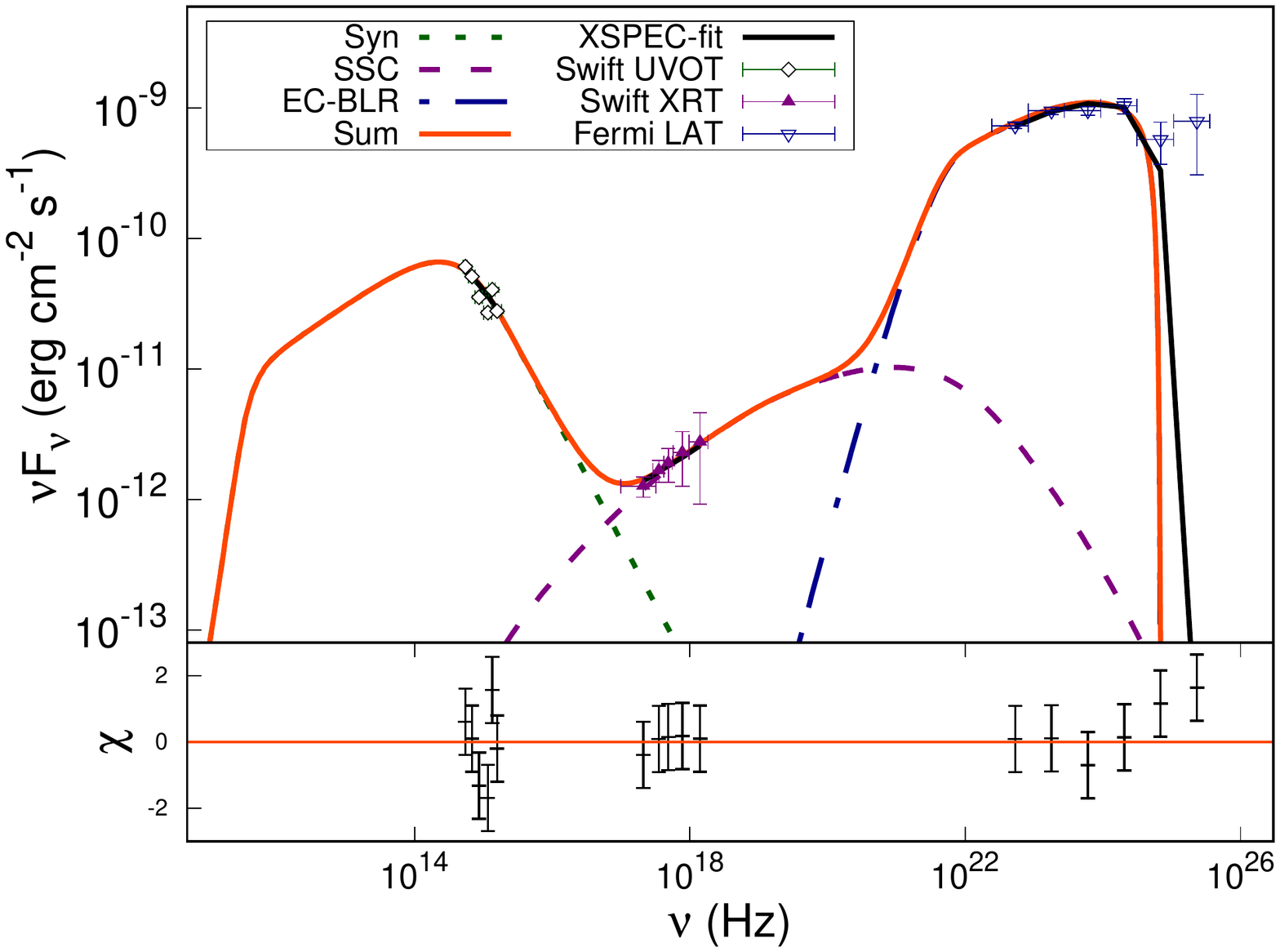}
\vspace*{-1cm}
        \caption{SED of \mbox{PKS\,0903-57} obtained during the flux state RFS. The observed flux points are specified by filled squares (\emph{Swift}-UVOT), open circles (\emph{Swift}-XRT), and filled circles (\emph{Fermi}-LAT). Left and right panels correspond to seed photons from IR torus and BLR photons, respectively. }
        \label{fig:sed_rfs}
		\end{center}        
\end{figure*}

\begin{figure*}
		\begin{center}
       \includegraphics[angle=0,width=.58\textwidth]{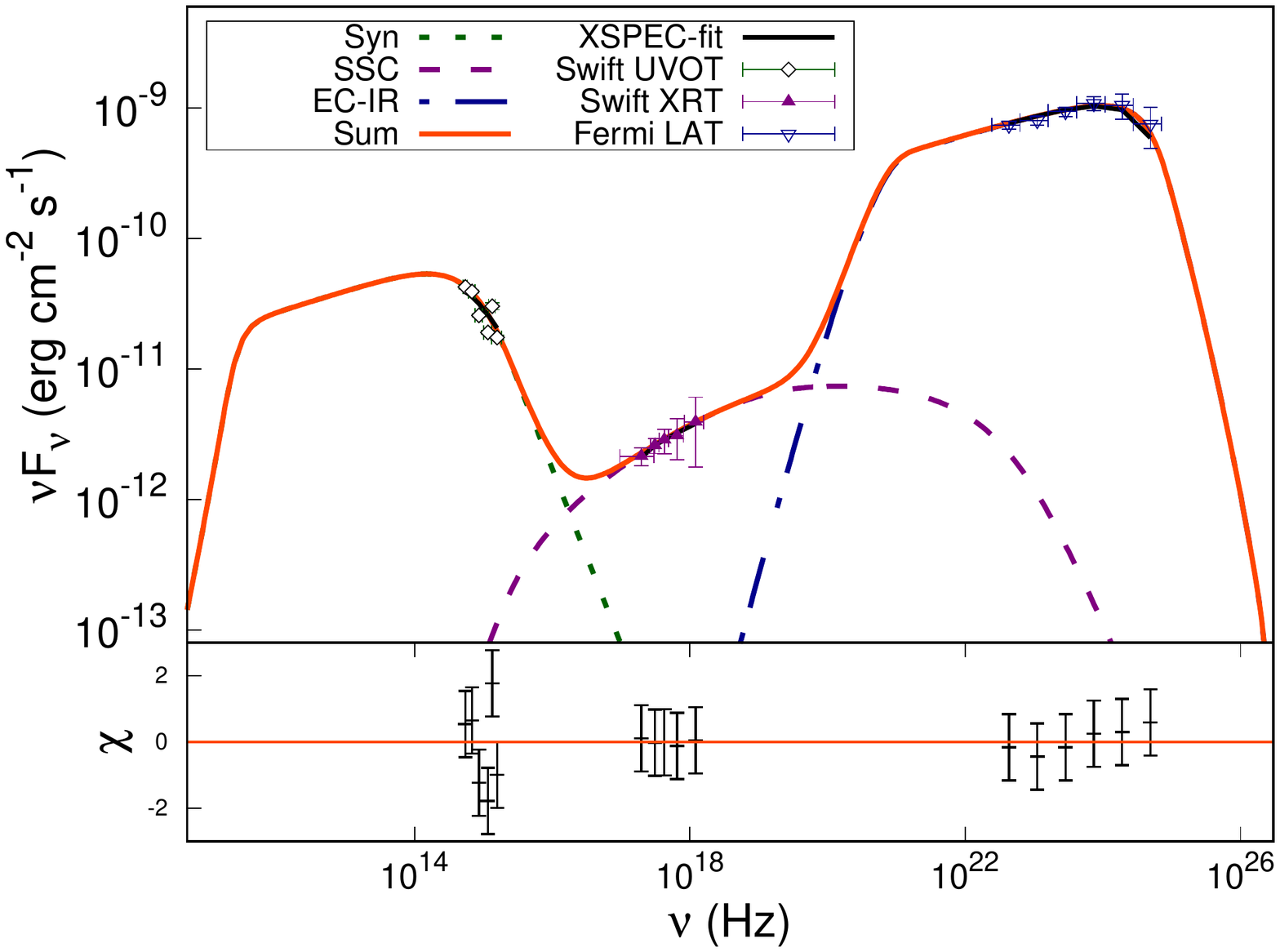}\hspace{-3.0cm}
        \includegraphics[angle=0,width=.58\textwidth]{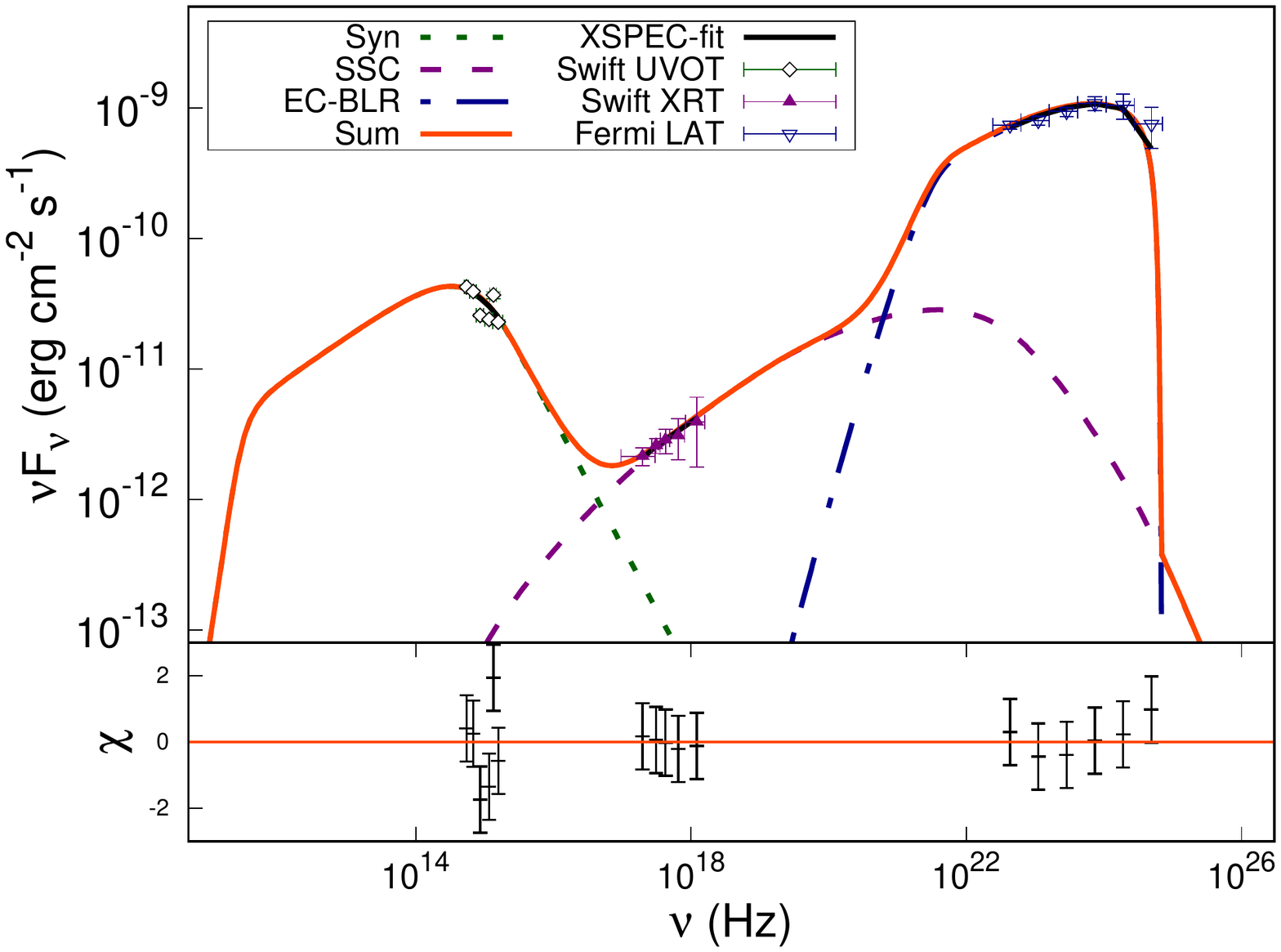}
	\vspace*{-1cm}    
        \caption{SED of \mbox{PKS\,0903-57} obtained during the flux state DFS. The labelling is same as that of Figure \ref{fig:sed_rfs}.} 
        \label{fig:sed_dfs}
		\end{center}        
\end{figure*}

\begin{figure*}
		\begin{center}
       \includegraphics[angle=0,width=.58\textwidth]{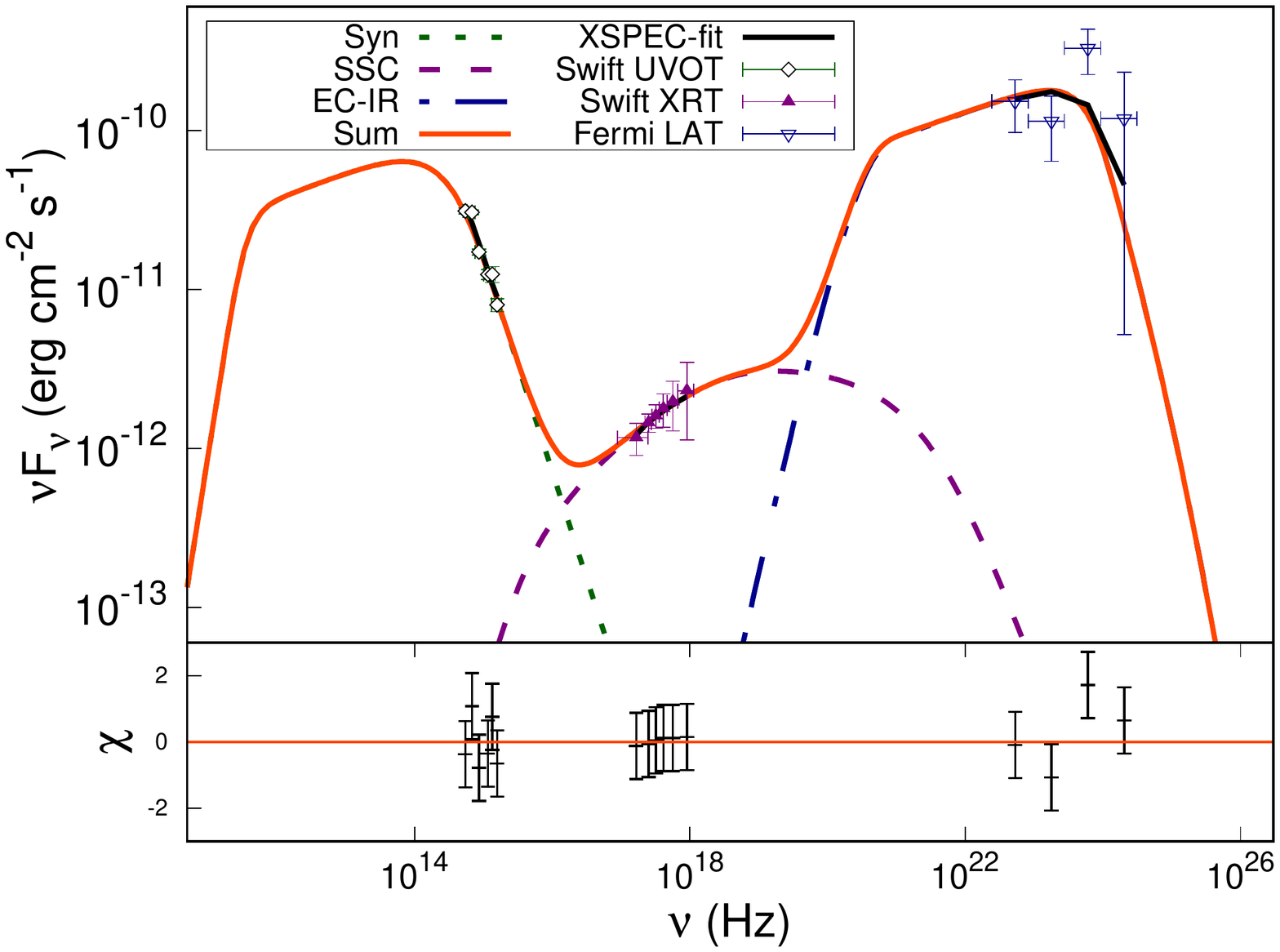}\hspace{-3.0cm}
        \includegraphics[angle=0,width=.58\textwidth]{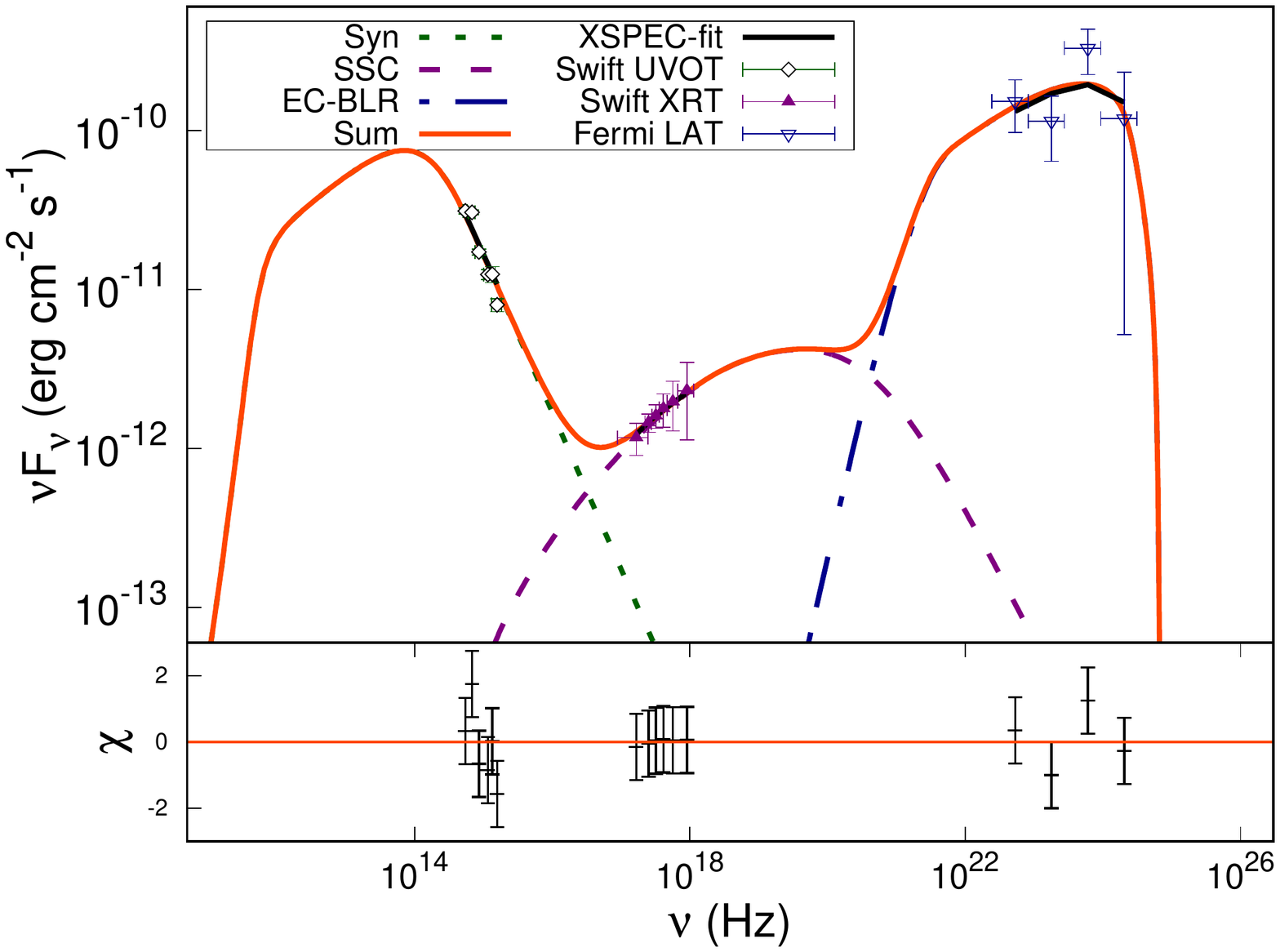}
	 \vspace*{-1cm}  
        \caption{SED of \mbox{PKS\,0903-57} obtained during the flux state QS1. The labelling is same as that of Figure \ref{fig:sed_rfs}.} 
        \label{fig:sed_qs1}
		\end{center}        
\end{figure*}

\begin{figure*}
		\begin{center}
       \includegraphics[angle=0,width=.58\textwidth]{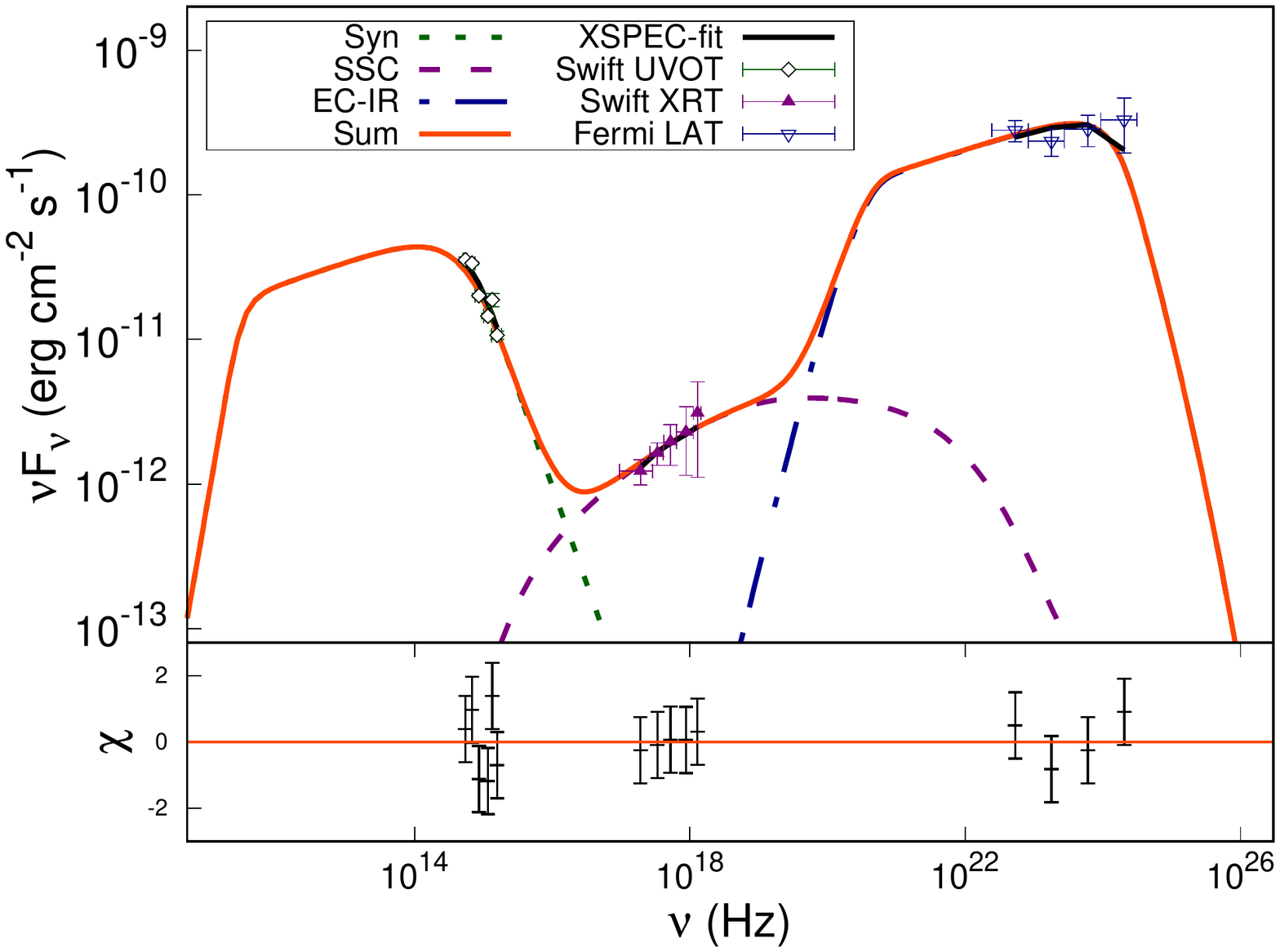}\hspace{-3.0cm}
        \includegraphics[angle=0,width=.58\textwidth]{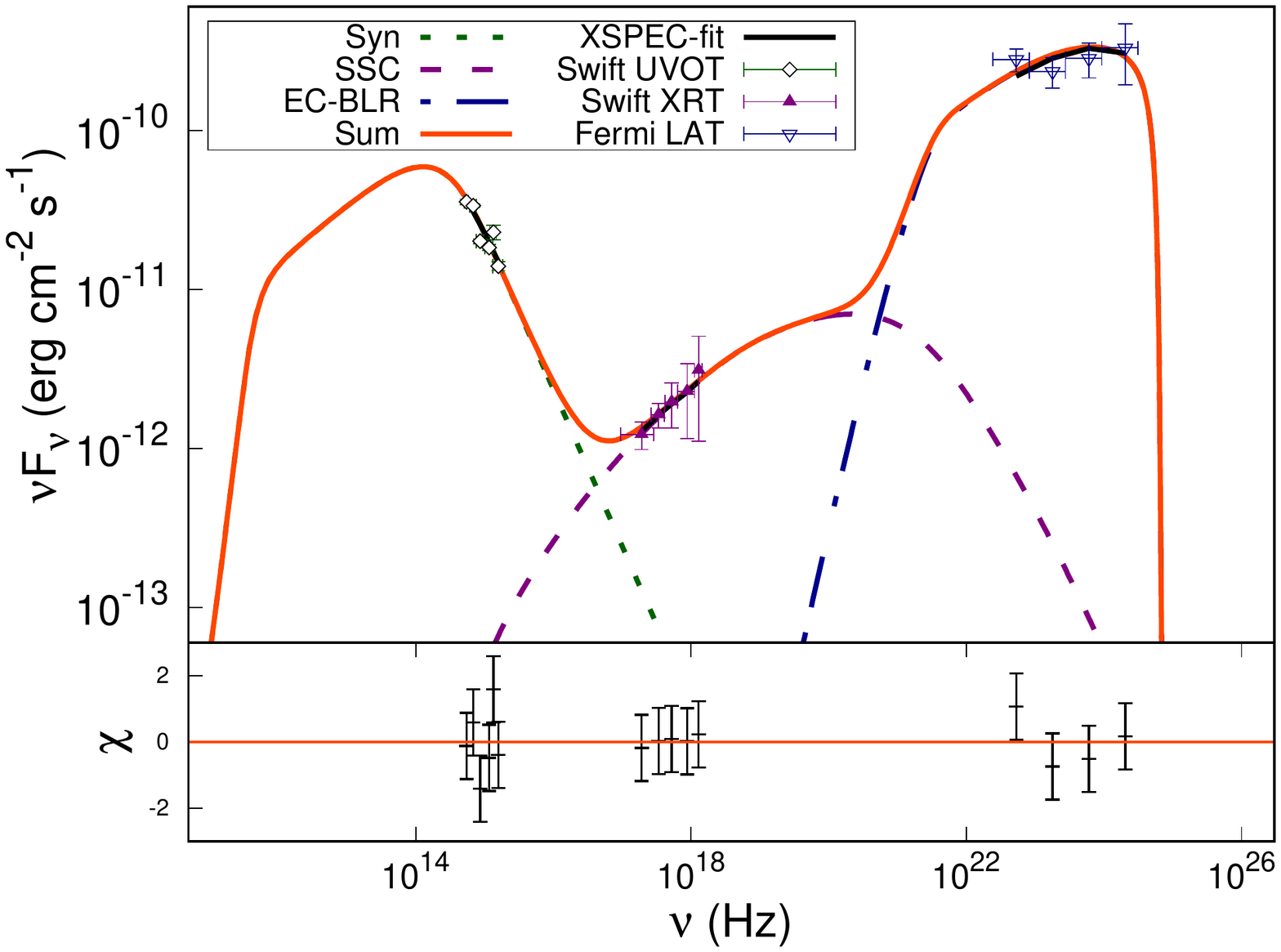}
	\vspace*{-1cm}    
        \caption{SED of \mbox{PKS\,0903-57} obtained during the flux state QS2. The labelling is same as that of Figure \ref{fig:sed_rfs}.}       
        \label{fig:sed_qs2}
		\end{center}        
\end{figure*}

\begin{figure*}
		\begin{center}
       \includegraphics[angle=0,width=.58\textwidth]{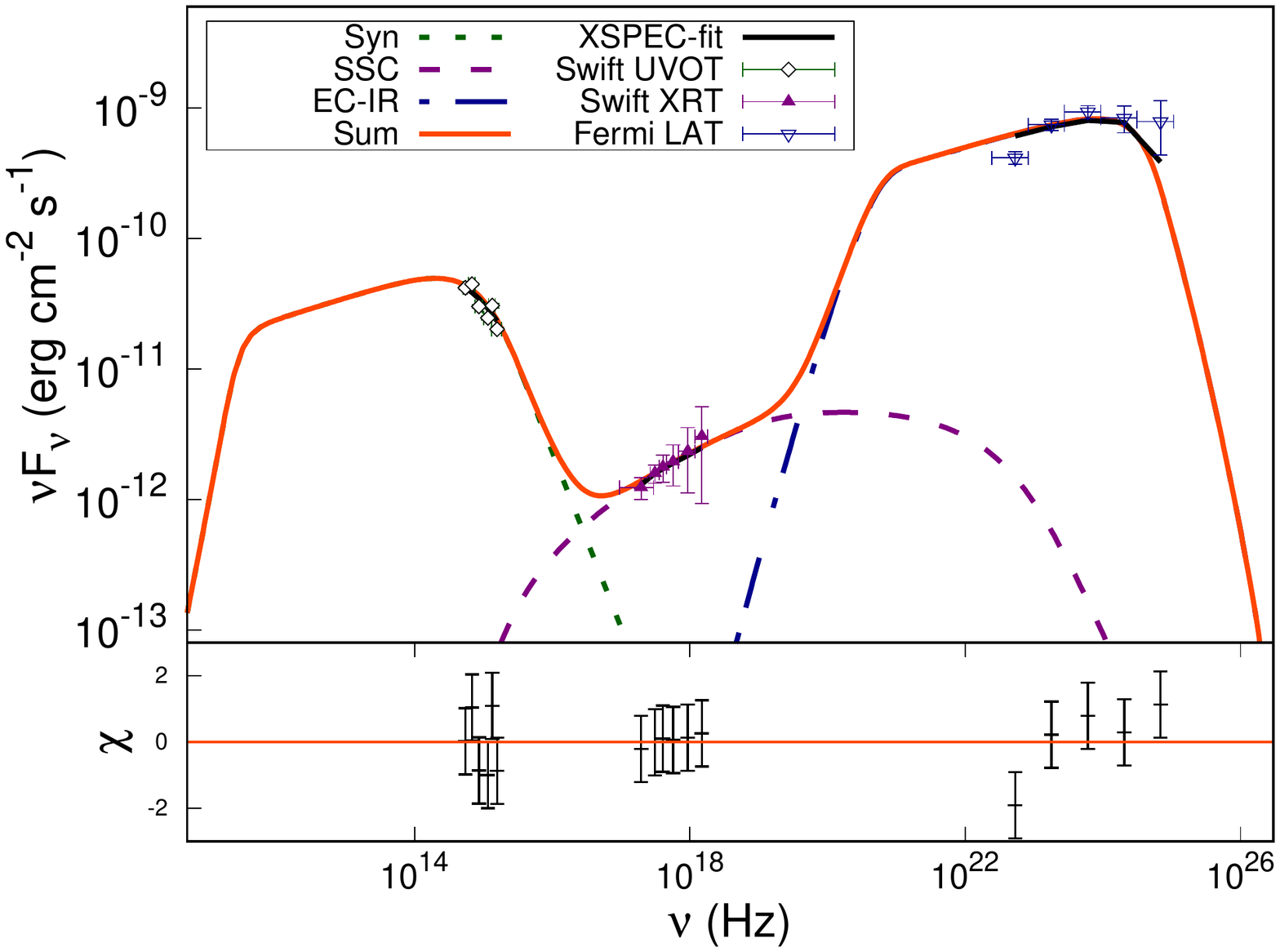}\hspace{-3.0cm}
        \includegraphics[angle=0,width=.58\textwidth]{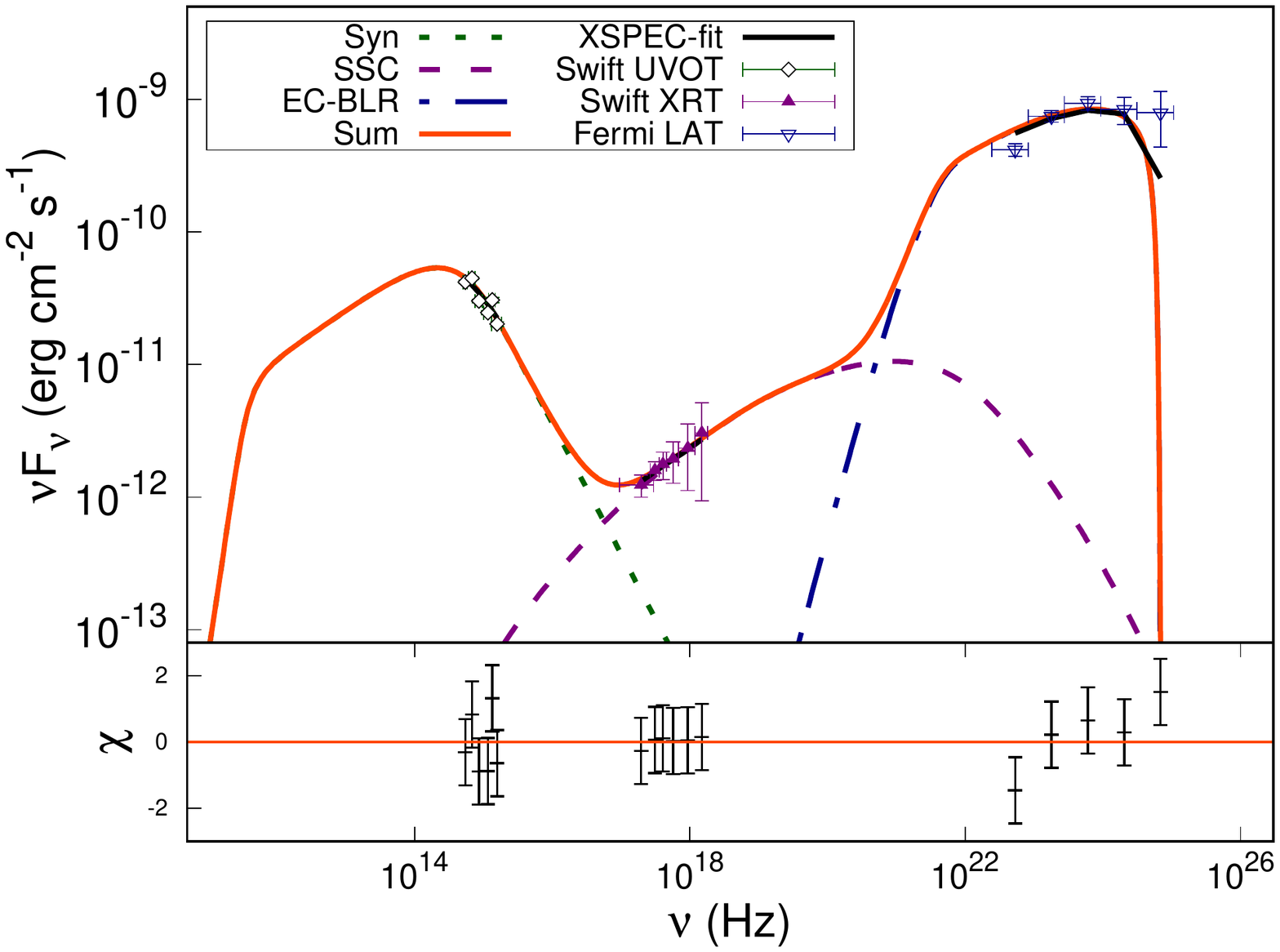}
	\vspace*{-1cm}    
        \caption{SED of \mbox{PKS\,0903-57} obtained during the flux state SVS. The labelling is same as that of Figure \ref{fig:sed_rfs}.} 
        \label{fig:sed_svs}
		\end{center}        
\end{figure*}

\begin{figure*}
		\begin{center}
       \includegraphics[angle=0,width=.58\textwidth]{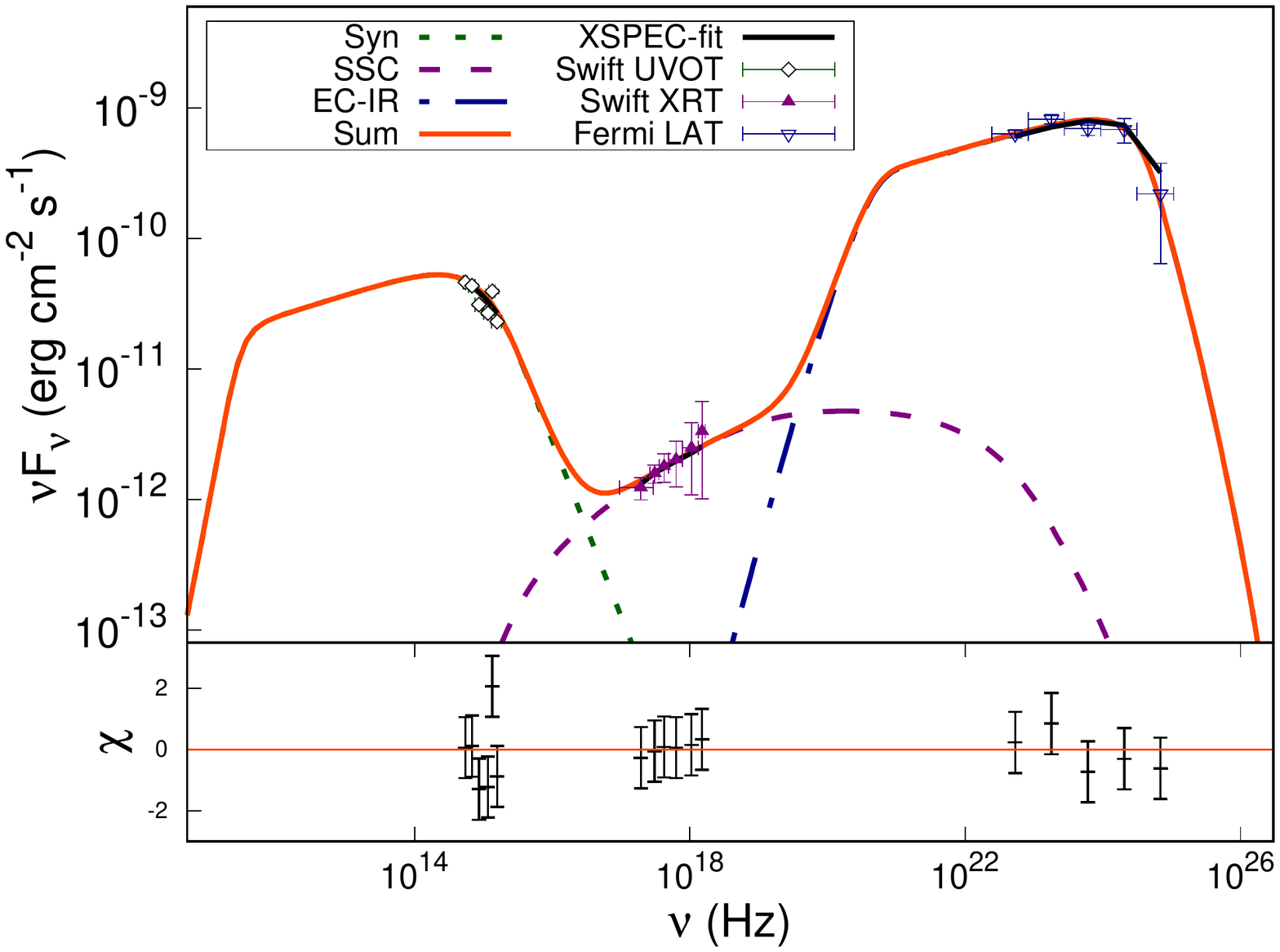}\hspace{-3.0cm}
        \includegraphics[angle=0,width=.58\textwidth]{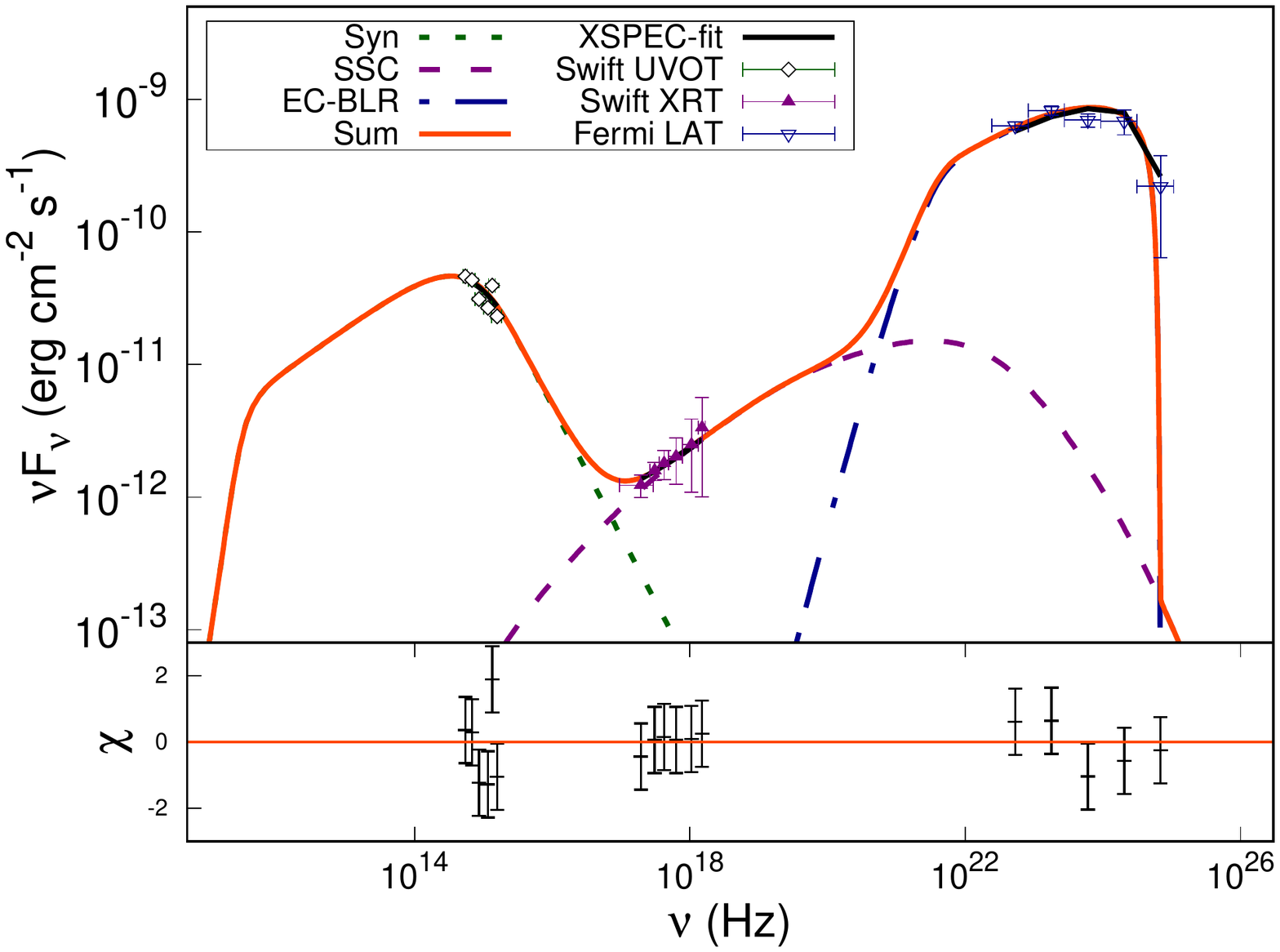}
	 \vspace*{-1cm}   
        \caption{SED of \mbox{PKS\,0903-57} obtained during the flux state VS. The labelling is same as that of Figure \ref{fig:sed_rfs}.}     
        \label{fig:sed_vs}
		\end{center}        
\end{figure*}

\subsection{Broadband Spectral Analysis}
In order to study the broadband spectral characteristics of \mbox{PKS\,0903-57} in different flux states, we chose six different time intervals from the multi-wavelength lightcurve based on: a) flare pattern and simultaneous observations available in $\gamma$-ray, X-ray and UV/optical energies,  b) good photon counts for reliable statistical fitting. These time intervals are shown by vertical lines in Figure \ref{fig:multiplot_lc} and are identified as Rising Flare State (RFS: MJD 58936--58940), Decaying Flare State (DFS: MJD 58943.2--58946.2), 
Quiescent State First (QS1: 58941.2--58942.0), Quiescent State Second (QS2: 58946.3--58947.7), Slow Varying State (SVS: 58951.8--58953.7) and Variable State (VS: 58954.1--58957). The integrated $\gamma$-ray spectrum in the considered flux states were modelled with power law 
and log-parabola model, $\frac{dN}{dE}=N_0(E/E_0)^{-\alpha-\beta\log(E/E_0)}$, where $N_0$ is normalisation, $\alpha$ is photon index at reference energy $E_0$, $\beta$ is curvature parameter, and the corresponding fitting parameters are given in Table \ref{table:spec_fit_param}. The spectral points for the broadband SED modelling are obtained from the log-parabola fit, if the curvature in the $\gamma$-ray spectrum is significant. In order to assess for the significance of curvature in the spectrum, we calculated the test statistics of the curvature, which is defined as $TS_{curve}=2[\log\mathcal{L}(LP)-\log\mathcal{L}(PL)$] \citep{2012ApJS..199...31N}. The curvature is considered to be significant, if the value of $TS_{curve}>16$. 
As shown in Table \ref{table:spec_fit_param}, significant curvature is observed in RFS ($TS_{curve}=16.24$), SVS ($TS_{curve}=16.12$) and VS ($TS_{curve}=17.84$).  
In each of considered flux states, the $\gamma$-ray SED points are obtained by dividing the total energy \mbox{0.1--500 GeV} into 8 energy bins equally spaced in log scale. During the spectral fitting in each energy bin, we assume the spectral parameters of sources other than \mbox{PKS\,0903-57} in the ROI does not change with energy, hence we froze the parameters of these sources to their best-fitting values obtained in the energy range \mbox{0.1--500 GeV}. The X-ray spectrum in each flux state is obtained by giving the specific observation ID to an online automated products generator tool \citep{2009MNRAS.397.1177E}. The source spectra is binned using the GRPPHA task in order to acquire 20 counts per bin and the resultant spectra is fitted with an absorbed power law model. In case of \emph{Swift}-UVOT, the images of the observation IDs which fall in particular flux state are combined using the UVOTIMSUM task and the optical/UV flux values are obtained from the combined image. The broadband SED points obtained in the RFS, DFS, QS1, QS2, SVS and VS are shown in Figures \ref{fig:sed_rfs}, \ref{fig:sed_dfs}, \ref{fig:sed_qs1}, \ref{fig:sed_qs2}, \ref{fig:sed_svs} and \ref{fig:sed_vs} respectively.

In order to model the broadband SED, corresponding to the different flux states considered, we assume  the emission to arise from  a spherical blob of characteristic radius `$R$'. The blob moves down the jet at relativistic speed with bulk- Lorentz factor $\Gamma$ at a small angle $\theta$ with respect the observer. The relativistic motion at small angle with respect to observer results in the Doppler boosting of observed flux, which is determined by beaming factor $\delta=1/\Gamma(1-\beta\cos\theta)$. If the duration of flare is controlled by the light travel time effect, then comoving radius of the emitting blob can be expressed as $R\approx\delta t_{var}/(1+z)$, where $t_{var}$ is the minimum observed variability timescale. We assume that the emission region to be populated with non-thermal electrons having broken power law  energy distribution 
\begin{align}
\label{eq:broken}
N(\gamma) d\gamma =\left\{
	\begin{array}{ll}
K \gamma^{-p}d\gamma,&\mbox {~$\gamma_{{\rm min}}<\gamma<\gamma_b$~} \\
K \gamma^{q-p}_b \gamma^{-q}d\gamma,&\mbox {~$\gamma_b<\gamma<\gamma_{{\rm max}}$~} 
\end{array}
\right.
\end{align}
where $\gamma$ is the dimensionless energy (electron Lorentz factor), $\gamma_b$ is the break energy, K is normalisation, p and q are the particle spectral index before and after break energy. The relativistic electrons 
emit radiation  through synchrotron, SSC and EC processes. The high energy (X-ray and $\gamma$-ray) emission in blazars is associated with the SSC and EC process. The seed photons for the EC process depends on the location of emission region and can be the photons from either BLR or IR torus. Hence, we consider two cases of seed photons for EC scattering i.e., seed photons from BLR and IR torus. These target photon fields are approximated to be isotropic blackbody with temperature $T_*$ and energy density $U_*=\frac{4f\sigma_{SB}}{c}\left(\frac{T_*}{2.82}\right)^4$, here f is the fraction of photons which gets IC scattered and $\sigma_{SB}$ is the Stefan-Boltzmann constant. The resultant spectra corresponding to synchrotron, SSC and EC processes are estimated numerically from their emissivity functions and the numerical code is incorporated as a local model in XSPEC to perform a statistical fitting of the broadband SEDs \citep{2018RAA....18...35S}.  In order to account for the model related uncertainties and systematic error in the data, we added a systematic of 15\% in the data. This is done by increasing the error bar of each data point by
adding in quadrature 15\% of the value of the energy flux data point. Using this model, the observed broadband spectrum can be reproduced mainly by 10 source parameters namely $\gamma_b$, p, q, $\Gamma$, B, R, $U_e$, $\theta$, f  and $T_*$ \citep{2018RAA....18...35S, 2019MNRAS.484.3168S}. We performed the fitting with $\gamma_b$, $\Gamma$, $U_e$ and B as free parameters, while other parameter were kept fixed to their typical values due to limited information that can be extracted from Optical/UV, X-ray and $\gamma$-ray observations. For numerical stability, we approximate the BLR emission to be a blackbody at temperature 42000 K (equivalent temperature corresponding to Lyman alpha line emission at $2.5\times 10^{15}$ Hz). The temperature of the IR photon is fixed at 900 K. 
Since $\gamma$-ray SED points in flaring state are comparatively better constrained than that of quiescent state SED points i.e., more upper-limits are in quiescent state than the flaring state, typical values of the fixed parameters viz. p, q, R and $\theta$ are obtained by fitting the RFS SED.  On considering the seed photons from the IR torus, the best-fit to the broadband SED of RFS is obtained by choosing the fixed parameters as $p=2.67$, $q=5.69$, $\rm R=5.21\times10^{16}\, cm$, $\rm\theta=0.1 \,degree$ and $f=3.19\times 10^{-3}$.
 The values of $\theta$ and R are chosen such that a near equipartition is satisfied and variability time scale is typically of the order of hours.  These typical values of fixed parameters are then used for fitting the broadband SEDs of DFS, QS1, QS2, SVS and VS. The best-fit spectral model consisting of synchrotron, SSC and EC components along with observed SED points for RFS, DFS, QS1, QS2, SVS and VS are shown in left panel of Figure \ref{fig:sed_rfs}, \ref{fig:sed_dfs}, \ref{fig:sed_qs1}, \ref{fig:sed_qs2},  \ref{fig:sed_svs} and \ref{fig:sed_vs} respectively, and the corresponding fitting parameters are given in Table \ref{table:sed}. Similarly, if we consider seed photons for EC scattering from BLR, then typical values for fixed parameters are obtained as $p=2.26$, $q=4.97$, $\rm R=1.88\times10^{16}cm$, $\rm\theta=0.1\,degree$ and $f=10^{-8}$. The resultant best-fit model SED along with observed points for the considered flux states are shown in the right panel of Figures \ref{fig:sed_rfs}, \ref{fig:sed_dfs}, \ref{fig:sed_qs1}, \ref{fig:sed_qs2}, \ref{fig:sed_svs} and \ref{fig:sed_vs}, and corresponding best fit parameters are given in Table \ref{table:sed}. Though both EC/IR and EC/BLR provides reasonable fit to the observed spectrum:  however, the reduced $\chi^2$ obtained in flux states viz. RFS, DFS, QS1 and VS indicates that EC/IR scattering provides comparatively good fit than EC/BLR scattering (see Table \ref{table:sed}). While in QS2 and SVS, the reduced-$\chi^2$ obtained from EC/BLR is comparatively smaller than EC/IR (see Table \ref{table:sed}). Further we found that the combination of seed photons for EC process from both BLR and IR torus (EC/(BLR+IR)) does not improve the fit in the considered flux states and instead the reduced $\chi^2$ were relatively larger than those obtained from EC/BLR and EC/IR. Hence, we did not include the analysis details for this case.

It should be noted that the one-zone leptonic model presented here assume that, in each of the time intervals probed, the electron energy distribution has achieved an equilibrium, and hence that, within each of these time intervals, the broadband emission is steady. However, as shown in Figure \ref{fig:daily_3h_rise_fal} and Figure \ref{fig:multiplot_lc}, there are substantial flux variations in the multi-band emission of \mbox{PKS\,0903-57} during these time intervals. Hence for the time intervals (e.g. RFS, DFS, VS), the electron energy distribution is actually far from equilibrium. In other words, the acquired multi-band data probes wrong one of the main assumptions required for the reliability of the SED model results presented here. In any case, for the exercise shown in this section, we neglect these multi-band flux variations, and consider only the averaged measured broadband SED in each of these time intervals. Therefore, the results shown here should be considered only as tentative, and are meant to highlight the variation of physical parameters among different time-averaged flux states.

\section{Summary and Discussion}\label{sec:discus}

The long-term flaring of \mbox{PKS\,0903-57} together with the availability of simultaneous observations from \emph{Fermi}-LAT and \emph{Swift}-XRT/UVOT has made possible for the first time to study the broadband temporal and spectral properties of \mbox{PKS\,0903-57} in detail. The continuous monitoring of \emph{Fermi}-LAT at $\gamma$-ray energy  reveals the presence of multiple flaring components in the active state (MJD 58931--58970) with the Comp-5 (fast-rise/slow-decay profile)  and  Comp-a (slow-rise/fast-decay profile)
showing a significant asymmetry, 
while Comp-4 (fast-rise/slow-decay profile), Comp-b (slow-rise/fast-decay profile) and Comp-c (slow-rise/fast-decay profile) are moderately asymmetric,
and Comp-1, Comp-2 and Comp-3 are symmetric. The flare profile can be due to the light travel time effects. Alternatively, the asymmetry in the flare profile can be attributed to the strengthening and weakening of acceleration process. A slow rise of the flare could be attributed to the acceleration of particles to higher energies while fast decay can be associated with the rapid loss of energy of high energy particles. 

 The analysis of the $\gamma$-ray spectrum in the considered flux states shows a significant curvature in RFS ($TS_{curve} = 16.24$) and  SVS ($TS_{curve}=16.12$) and VS ($TS_{curve}=17.84$).  Since a power-law distribution of emitting electrons produces a power-law spectrum, using the same analogy, the observed curved spectra indicates that the underlying emitting electron distribution is curved. 
The standard approach used to explain the curvature in the spectrum is to consider combined effects of particle acceleration and radiative losses. For example, \citet{2004A&A...413..489M} showed that the curved spectrum  can be produced through the energy dependent acceleration process such that the acceleration probability is a decreasing function of the electron energy.
Alternatively, an energy dependent electron escape rate from the acceleration region can also result in the curvature of the emitted spectrum \citep{2018MNRAS.478L.105J, 2018MNRAS.480.2046G}.

The temporal analysis of MLC suggests that the source exhibits significant variability at all energy bands, indicating a single emission region and same electron population may be responsible for the flaring emission. Observation of large variability amplitude at $\gamma$-ray energy than the X-ray and optical/UV bands is consistent with other blazars \citep[e.g. ][]{2005ApJ...629..686Z, 2019MNRAS.484.3168S}. The large amplitude variations at $\gamma$-rays suggest that $\gamma$-ray emission is due to high energy electrons, while small variations at low photon energy may arise from the low energy  electron distribution. This is due to the fact that the high-energy electrons cool much faster than low-energy electrons. 
The increase in variability amplitude with the increase in photon energy can also reveal the signature of spectral variability of the source \citep{2005ApJ...629..686Z}.

The broadband SED of \mbox{PKS\,0903-57} in different flux states are reproduced by considering the synchrotron, SSC and EC processes. Here we assume that the electron energy distribution in each of the flux states  has reached an equilibrium, such that the broadband emission is steady in these time intervals. However, the substantial flux variations in the multi-band emission of \mbox{PKS\,0903-57} during these time intervals (see Figure \ref{fig:daily_3h_rise_fal} and \ref{fig:multiplot_lc}) indicates that the electron energy distribution is actually far from equilibrium in the considered time intervals (e.g., RFS, DFS and VS). This implies that the acquired multi-band data actually goes against the assumptions required for the reliability of the SED model results. In our study, we neglect these multi-band flux variations, and consider only the averaged measured broadband SED in each of these time intervals. Under this scenario, the target photons from the IR torus (EC/IR) or BLR-region (EC/BLR) provides acceptable fit to all flux states, however, the reduced $\chi^2$ obtained
in higher flux states viz. RFS, DFS and VS  for EC/IR scattering are slightly better than those obtained in EC/BLR (see Table \ref{table:sed}).
 During the active state, several HE events ($\rm \geq 10\,GeV$) were detected with 99\% or more probability of being associated with the source direction. The detection of HE photons ($\rm \geq 10\,GeV$) suggest that the emission region is located out side the BLR region in order to avoid pair absorption in collisions with low-energy photons \citep{2006ApJ...653.1089L}. Additionally, the ratio of $U_e/U_B$ is more close to equipartition value in case of EC/IR model than those obtained in EC/BLR. Therefore, detection of  HE photons ($\rm \geq 10\,GeV$) and equipartition values suggest that the $\gamma$-ray emission is more likely to be produced due to EC scattering of IR photons. Moreover, the Klein–Nishina effect steepens the EC/BLR spectrum  sharply at much lower energy ($\sim 50$ GeV) than in case of EC/IR spectrum, which makes the VHE detection difficult in EC/BLR scattering. Therefore, detection of VHE emission from \mbox{PKS\,0903-57} during the active state \citep{2020ATel13632....1W} further supports EC/IR mechanism and hence the
location of the emission region may be outside the BLR region. 
On considering the EC/IR scattering, the rise in flux is mainly associated with the increase in $\Gamma$ and $\gamma_b$, and decrease in the magnetic field (see Table \ref{table:sed}). In addition to these, the change in particle energy density are also noticed in the considered flux states. Moreover, on comparing the SED of quiescent and high flux states (RFS, DFS, SVS and VS), we noticed that flux enhancement at $\gamma$-ray energy is larger than at lower energy bands. The rise in integrated $\gamma$-ray flux (\mbox{0.1--500 GeV}) from QS1 to RFS is by a factor of $\sim 5$, while in optical/UV, the maximum increase in flux from QS1 to RFS is obtained in filter UVW2, it is only by a factor of $\sim 3.5$. Since the $\gamma$-ray emission is associated with the IC scattering of external target photon field, the flux increase at $\gamma$-ray energy can be attributed to the increase in bulk Lorentz factor of the emission region which enhances the target photon energy density by $\Gamma^2$ in the rest frame of emission region.

We estimate the jet power by assuming that the protons are relatively cold with their
number density equal to that of non-thermal electrons. The cold protons provide 
inertia to the jet and the non-thermal electrons are responsible for the radiation.
The total jet power in terms of proton energy density, electron energy density, and the magnetic field energy density can be written as \citep{2008MNRAS.385..283C}
\begin{equation}
P_{jet}=\pi R^2\Gamma^2\beta_{\Gamma}c(U_p+U_e+U_B)
\end{equation}
The $ P_{jet}$ for different flux states along with the total radiated power
obtained are given in Table 4. For the chosen set of parameters, we find that the radiated power at the blazar zone of the jet is much less than the kinetic power. This implies that only minimum amount of the jet power is spent at the blazar zone and most of the energy is retained to launch the jet up to kpc/Mpc scales.

The cooling time of electrons emitting $\gamma$-ray photons with energy ($\epsilon_{\gamma}$) through the IC scattering of IR photons ($\epsilon_0$) can be obtained using the relation
\begin{equation}
\tau_{cool}\approx \frac{3m_ec}{4\sigma_T U'_{tot}}\sqrt{\frac{\Gamma\epsilon_0(1+z)}{\delta\epsilon_{\gamma}}}\quad,
\end{equation} 
where $U'_{tot}=U_B+U_{syn}+U_{IR}$ is energy density of magnetic field $U_B$, synchrotron photons $U_{syn}$ and seed photons from IR region $U_{IR}$ in the emission region frame. For seed photon temperature of 900 K and $\epsilon_{\gamma}= 0.2$ GeV,  the cooling time obtained for RFS and DFS are $\sim 9$ hr and $\sim 12$ hr, respectively, which are smaller than the flare rise/fall times in Comp-1 (which includes RFS) and Comp-2 (which includes DFS) respectively (see Table  \ref{table:multi-comp}). 
The cooling time of electrons emitting $\gamma$-ray radiation is  also reported shorter than the  flare time in other blazars \citep[e.g.,][]{2013ApJ...766L..11S, 2015ApJ...804...74P}.  The cooling timescale shorter than flare rise/decay times (see Table \ref{table:multi-comp}) indicates that the flare profile is possibly governed by the light crossing time. 
This also implies that the flare emission may be influenced by factors like geometry or  in-homogeneous electron density of the emitting region \citep{2009ApJ...692.1374B, 2014MNRAS.442..131K}. On the other hand,  \citet{2016ApJ...819..156B} showed that cooling timescales shorter than variability timescales can be better explained by considering the emission being produced by multiple zones.

The observed broadband SED of \mbox{PKS\,0903-57} in different flux states shows the Compton dominance. The peak flux of high energy components is by a factor $> 10$ larger than low energy component (see Figure \ref{fig:sed_rfs}, \ref{fig:sed_dfs}, \ref{fig:sed_svs} and \ref{fig:sed_vs}). Among the blazar subclass, FSRQ shows Compton dominance \citep{2010AIPC.1223...79T} and this require both SSC and EC process to reproduce high energy  (X-ray and $\gamma$-ray) emission \citep{2017MNRAS.470.3283S,2018RAA....18...35S,2019MNRAS.484.3168S}.  While in the low state of HBLs, the high energy emission can be explained by SSC process  alone \citep{2011ApJ...736..131A, 2011ApJ...733...14M, 2014MNRAS.439.2933Y}, however during flaring, the broadband SED sometimes show sub-structures whose explanation require a multi-zone scenario, or even more ``exotic" models, as reported for Mrk\,501 \citep{2020A&A...637A..86M}.  Therefore in \mbox{PKS\,0903-57}, the observation of Compton dominance and the requirement of  SSC and EC process to produce high energy  emission indicates that \mbox{PKS\,0903-57} is more likely a FSRQ/LBL type blazar. 
The classification can be further confirmed by performing a detailed optical spectroscopic observation of the source.

\section{ACKNOWLEDGEMENTS}
We thank the anonymous referee for the useful comments and suggestions which improved the manuscript. ZS thanks Ranjeev Misra for useful comments, discussions and helpful support.
This research has made use of $\gamma$-ray data from Fermi Science Support Center (FSSC). The work has also used the \emph{Swift} Data from the High Energy Astrophysics Science Archive Research Center (HEASARC), at NASA's Goddard Space Flight Center. 

\section{Data availability}
The data and the model used in this article will be shared on reasonable request to the corresponding author, Zahir Shah (email: zahir@iucaa.in or shahzahir4@gmail.com).

\bibliographystyle{mnras}
\bibliography{references} 

\bsp	
\label{lastpage}
\end{document}